\documentclass[11pt]{article}

\usepackage{epsfig, psfrag}
\usepackage{latexsym}
\usepackage{enumerate}
\usepackage{url}
\usepackage{float}
\usepackage[hypertexnames=false,hyperfootnotes=false]{hyperref}
\usepackage{color}
\usepackage{tikz}
\usepackage[margin=10pt,font=small,labelfont=bf]{caption}
\usepackage[subrefformat=parens,labelformat=simple]{subcaption}
\usepackage{afterpage}
\usepackage{enumitem}
\usepackage[boxed]{algorithm}
\usepackage{algpseudocode}
\usepackage[normalem]{ulem}
\usepackage{lmodern}
\usepackage{booktabs}
\usepackage{sectsty}
\usepackage{ifthen}
\usepackage{amsmath}
\usepackage{amsthm}
\usepackage{amssymb}
\usepackage{amsfonts}
\usepackage{thmtools}
\usepackage{thm-restate}
\usepackage{xspace}
\usepackage{titling}
\usepackage{parskip}
\usepackage{natbib}
\usepackage{xfrac}
\usepackage{multirow}
\usepackage{bigdelim}
\usepackage{bm}
\newcommand{\E}{\ensuremath{\mathsf{E}}} % expectation

\DeclareMathOperator{\diag}{diag}

\DeclareMathOperator{\Var}{Var}
\DeclareMathOperator{\Cov}{Cov}

\DeclareMathOperator*{\argmin}{\mathrm{argmin}}
\DeclareMathOperator*{\argmax}{\mathrm{argmax}}

\declaretheoremstyle[headfont=\sffamily\bfseries,bodyfont=\itshape]{thm-sf}

\declaretheorem[style=thm-sf]{remark}

\declaretheorem[style=thm-sf]{corollary}

\declaretheorem[style=thm-sf]{proposition}

\newcommand{\proofnamest}[1]{{\normalfont\sffamily\bfseries #1}}
\renewcommand{\thmcontinues}[1]{\hyperref[#1]{continued}}
\usetikzlibrary{arrows,patterns,plotmarks,pgfplots.groupplots}
\tikzstyle{every picture} += [>=stealth]
\tikzset{axis/.style={semithick, line join=miter}}
\allsectionsfont{\sffamily}
\makeatletter
\def\@seccntformat#1{\csname the#1\endcsname.\quad}
\makeatother
\floatname{algorithm}{\normalfont\sffamily\bfseries Algorithm}
\floatstyle{ruled}
\provideboolean{fastcompile}
\newcommand{\hidefastcompile}[1]{\ifthenelse{\boolean{fastcompile}}{}{#1}}
\newcommand{\todo}[1]{{\color{red} \noindent {\sffamily\bfseries TODO:} #1}}

\usepackage{pgfplots}
\usepackage{pgfplotstable}
\usetikzlibrary{calc}
\usepackage{mathtools}
\definecolor{orange}{rgb}{0.85,0.33,0.13} % 217,85,33
\definecolor{green}{rgb}{0.13,0.85,0.33}
\definecolor{purple}{rgb}{0.33,0.13,0.85}
\definecolor{lime}{rgb}{0.65,0.85,0.13}
\definecolor{blue}{rgb}{0.13,0.65,0.85}
\pgfplotscreateplotcyclelist{tricolor}{%
  orange,every mark/.append style={fill=orange!80!black},mark=*\\%
  green,every mark/.append style={fill=green!80!black},mark=square*\\%
  purple,every mark/.append style={fill=purple!80!black},mark=otimes*\\%
  black,mark=star\\%
  orange,every mark/.append style={fill=orange!80!black},mark=diamond*\\%
  green,densely dashed,every mark/.append style={solid,fill=green!80!black},mark=*\\%
  purple,densely dashed,every mark/.append style={solid,fill=purple!80!black},mark=square*\\%
  black,densely dashed,every mark/.append style={solid,fill=gray},mark=otimes*\\%
  orange,densely dashed,mark=star,every mark/.append style=solid\\%
  green,densely dashed,every mark/.append style={solid,fill=green!80!black},mark=diamond*\\%
}
\pgfplotsset{colormap={tricolormap}{color=(orange) color=(green) color=(purple)},
  colormap={quadcolormap}{color=(orange) color=(lime) color=(blue) color=(purple)}}
\pgfplotstableset{%
  font=\small,
  every head row/.style={before row=\toprule[1pt], after row=\midrule},
  every last row/.style={after row=\bottomrule[1pt]}}
\usepackage{setspace}
\usepackage[margin=1in]{geometry}

\setcitestyle{authoryear,round,semicolon,aysep={,},yysep={,},notesep={, }}
\provideboolean{submissionversion}
\ifthenelse{\boolean{submissionversion}}{
  \renewcommand{\todo}[1]{}
  
  \newcommand{\deledit}[1]{}
}{
  
  \newcommand{\deledit}[1]{{\color{orange} \sout{#1}}}
}

\newcommand{\idio}{\textup{id}}
\newcommand{\fund}{\textup{f}}
\newcommand{\tran}{\textup{tr}}

\newcommand{\Pid}{\ensuremath{\bm{\bar{\Psi}}_\idio}}
\newcommand{\Pf}{\ensuremath{\bm{\bar{\Psi}}_\fund}}
\newcommand{\W}{\ensuremath{\mathbf{W}}}

\newcommand{\Pidt}{\ensuremath{\bm{\tilde{\Psi}}_\idio}}
\newcommand{\Pft}{\ensuremath{\bm{\tilde{\Psi}}_\fund}}
\newcommand{\Wt}{\ensuremath{\mathbf{\tilde{W}}}}

\tikzstyle{rate} += [color=orange,very thick]
\usetikzlibrary{matrix,positioning}
\usepgfplotslibrary{groupplots}
\pgfplotsset{compat=newest}

\ifthenelse{\boolean{submissionversion}}{
  \title{\textsf{\textbf{Cross-Sectional Variation of Intraday Liquidity, Cross-Impact, and their Effect on Portfolio Execution}}}
  \author{}
  \date{}
}{
\title{\textsf{\textbf{Cross-Sectional Variation of Intraday Liquidity, Cross-Impact, and their Effect on Portfolio Execution\thanks{
Graduate School of Business, Columbia University;
email: \texttt{\{smin20, c.maglaras, ciamac\}@gsb.columbia.edu}}
}}}
\author{ \\
  Seungki Min \\
  \and \\
  Costis Maglaras \\
  \and \\
  Ciamac C. Moallemi \\
}

\date{Initial Version: July 2017; December 2017 \\
 Current Revision: \today }
}

\begin{document} 

\maketitle
\singlespacing

\begin{abstract}
The composition of natural liquidity has been changing over time.
An analysis of intraday volumes for the S\&P500 constituent stocks illustrates that (i) volume surprises, i.e., deviations from their respective forecasts, are correlated across stocks, and (ii) this correlation increases during the last few hours of the trading session. 
These observations could be attributed, in part, to the prevalence of portfolio trading activity that is implicit in the growth of ETF, passive and systematic investment strategies; and, to the increased trading intensity of such strategies towards the end of the trading session, e.g., due to execution of mutual fund  inflows/outflows that are benchmarked to the closing price on each day.

In this paper, we investigate the consequences of such portfolio liquidity on price impact and portfolio execution. 
We derive a linear cross-asset market impact from a stylized model that explicitly captures the fact that a certain fraction of natural liquidity providers only trade portfolios of stocks whenever they choose to execute. 
We find that due to cross-impact and its intraday variation, it is optimal for a risk-neutral, cost minimizing liquidator to execute a portfolio of orders in a coupled manner, as opposed to a separable VWAP-like execution that is often assumed. 
The optimal schedule couples the execution of the various orders so as to be able to take advantage of increased portfolio liquidity towards the end of the day. 
A worst case analysis shows that the potential cost reduction from this  optimized execution schedule over the separable approach can be as high as 6\% for plausible model parameters. 
Finally, we discuss how to estimate cross-sectional price impact if one had a dataset of realized portfolio transaction records that exploits the low-rank structure of its coefficient matrix suggested by our analysis.
\end{abstract}

%\begin{center}
%  \bfseries\sffamily PRELIMINARY VERSION --- DO NOT DISTRIBUTE
%\end{center}

\onehalfspacing

\section{Introduction}

Throughout the past decade or so we have experienced a so-called movement of assets under management in the equities markets from actively managed to passively and systematically managed strategies.
This migration of assets has also been accompanied by the simultaneous growth of Exchange-Traded-Funds (ETFs). 
In very broad strokes, such strategies tend to make investment and trade decisions based on systematic portfolio-level procedures -- e.g., invest in all S\&P500 constituents proportionally to their respective market capitalization weights; invest in low-volatility stocks; high-beta stocks; high dividend stocks; etc. 
In contrast, active strategies, for example, may focus on fundamental analysis of individual firms that may, in turn, result into discretionary investment decisions on the respective stocks. 
In the sequel, we will refer to passive strategies as ``index fund'' strategies.

This gradual shift in investment styles has affected the nature of trade order flows, which motivates our subsequent analysis. 
We make three specific observations.
First, passive and systematic strategies tend to generate portfolio trade order flows, i.e., trades that simultaneously execute orders in multiple securities in a coordinated fashion; e.g., buying a \$50 million slice of the S\&P500 over the next 2 hrs that involves the simultaneous execution of buy orders along most or all of the index constituents.
Second, passive strategies tend to concentrate their trading activity towards the end of the day; in part, so as to focus around times with increased market liquidity, and because mutual funds that implement such strategies have to settle buy and sell trade instructions from their (retail) investors at the closing market price at the end of each day; ETF products exhibit similar behavior.
Third, the shift in asset ownership over time and the changes in the regulatory environment, have, in turn, changed the composition and strategies under which natural liquidity is provided in the market -- these are the counterparties that step in to either sell of buy stock against institutional investor interest so as to clear the market.
%an institutional investor, say interested to buy 1 million shares of MSFT, may eventually trade against; these could be risk trading desks in broker dealers that commit capital to facilitate the trade, or opportunistic investors that may choose to supply liquidity in response to a realized change in market prices. Continuing with the above example, index fund investors would typically not have the discretion to sell a large block of MFST stock without doing some contingent trading in other securities that are are part of the underlying systematic investment strategy, thus changing the nature of opportunistic liquidity provision; index fund investors may be capable, however, to provide cheaper portfolio liquidity in a direction that is well aligned with their investment strategy; e.g., sell a slice of a technology sector portfolio, as opposed to only one security.

In \S \ref{s-empirical} we will provide some empirical evidence that show that pairwise correlations amongst trading volumes across the S\&P500 constituents are positive throughout the trading day, and increase by about a factor of two over the last 1-2 hrs of the trading day.
That is, trading volumes exhibit common intraday variation away from their deterministic forecast in a way that is consistent with our earlier observations.

%Statistically, this observation basically confirms that intraday liquidity exhibits common variation plausibly due to the common order flows from ETFs and passive funds.

In this paper we study the effect of portfolio liquidity provision in the context of  optimal trade execution.
Specifically, we consider a stylized model of natural liquidity provision that incorporates the behavior of single stock and portfolio participants, and, in turn, leads to a market impact model that incorporates cross-security impact terms; these arise due to the participation of natural portfolio liquidity providers.
We formulate and solve a multi-period optimization problem to minimize the expected market impact cost incurred by a risk-neutral investor that seeks to liquidate a portfolio. 
We characterize the optimal policy, which is coupled -- i.e., the liquidation schedules for the various orders in the portfolio should be jointly determined so as to incorporate and exploit the cross-security impact phenomena. 
We contrast the optimal schedule against that of a separable execution approach, where the orders in the portfolio are executed independently of each other; 
this is commonly adopted by risk-neutral investors.
%which would indeed be optimal in the absence of cross-impact, or when single-stock and portfolio liquidity provision follow the same intraday intensity profiles/
Separable execution is suboptimal, in general, and we derive a bound on it sub-optimality gap when compared to the optimal (coupled) solution, which can be interpreted as the execution cost reduction that an investor can achieve by optimizing around such cross-impact effects.

In a bit more detail, the main contributions of the paper are the following.

\textbf{Stylized model of cross-sectional price impact:} 
Under the assumption that the magnitude of single stock and portfolio liquidity provision is linear in the change in short-term trading prices, we show that market impact is itself linear in the trade quantity vector, and characterize the  coefficient matrix that exhibits an intuitive structure:
it is the inverse of a matrix that is a diagonal --capturing the effect of single-stock liquidity providers -- plus a (non-diagonal) low-rank matrix --capturing the effect of portfolio liquidity providers that are assumed to trade along a set of portfolio weight vectors, such as the market and sector portfolios. 
Cross-impact is the result of portfolio liquidity provision.

The linear market impact model results in quadratic trading costs, which will allow for a tractable downstream analysis.
The derivation of this model suggests that cross-impact effects will also arise in settings where liquidity provision follows more complex and possibly non-linear strategies, as long as a portion of that liquidity is provided by portfolio investors.

\textbf{Optimal trade scheduling for risk-neutral minimum cost liquidation:} 
We formulate and solve an optimal multi-period optimization problem that selects the quantities to be traded in each security over time so as to liquidate the target portfolio over the span of a finite horizon (a day in our case) in a way that minimizes the cumulative expected market impact costs.
Coupling is not the result of a risk penalty that captures the covariance of intraday price returns, as is typically the case, but the result of correlated liquidity.
The optimal trade schedule is coupled, and, specifically, incorporates and exploits the presence and intraday variation of cross-impact effects.
We identify the special cases where a separable execution approach would be optimal, namely: a) if there was no portfolio liquidity provision, or b) the intensity of portfolio liquidity provision was constant throughout the trading day.
%Both of these cases seem not to agree with  

\textbf{Worst case analysis:} 
We compare the optimal policy against a separable (VWAP-like) execution approach, and characterize the worst case liquidation portfolios and the magnitude of the benefit that one derives from the optimized solution.
A straightforward estimation of the mixture of single-stock and portfolio liquidity providers that would be consistent with the intraday volume profile and the intraday profile of pairwise volume correlations, allows us to back into a numerical value for the aforementioned bound, which is 6\%.
The worst case analysis provides some intuition on the settings where this effects may be more pronounced. 

\textbf{Efficient estimation of cross-impact:} 
We suggest a practical scheme for estimating the (time-varying) coefficient matrix for price impact. 
A direct estimation procedure for all cross-impact coefficients between each pair of stocks seems intractable due to the low signal-to-noise ratio that often characterizes market impact model estimation, and the increased sparsity of trade data when we study pairs of stocks.
Exploiting the low rank structure of our stylized impact model derived above, we propose a procedure that only involves the estimation of a few  parameters, e.g., one parameter per sector. 
We do not calibrate the cross-impact model, as this typically requires access to proprietary trading information, but we specify a tractable maximum likelihood procedure that one could make use.

\subsection{Literature Survey} 
One set of papers that is related with our work focuses on optimal trade scheduling, where the investor considers a dynamic control problem for how to split the liquidation of a large order over a predetermined time horizon so as to optimize some performance criterion.
\cite{BertsimasLo98} solved this problem in the context of minimizing the expected market impact cost, 
and  \cite{AlmgrenChriss00} extended the analysis to the mean-variance criterion; see also \cite{Almgren03} and \cite{HubermanStanzl05}.
%Schied and Schoneborn considered an expected utility criterion.
%These papers use a linear impact model to capture the cost of the execution that does not incorporate cross-security impact.
\cite{BertsimasLo98} shows that the cost minimizing solution under a linear impact model schedules each order in proportion to the stock's forecasted volume profile.
In these papers, multiple-security trading is shortly discussed as an extension of single-stock execution, and the similar setup can be found in recent studies (e.g., \cite{brown10}, \cite{haugh14}).
A separate strand of work, which includes \cite{obizhaeva}, \cite{rosu} and \cite{alfonsi}, treat the market as one limit order book and use an aggregated and stylized model of market impact to capture how the price moves as a function of trading intensity.
% also abstracting away queueing effects.
 \cite{Tsoukalas17} builds on \cite{obizhaeva} to consider a portfolio liquidation problem incorporating risk and cross-impact effects (that shift the bid/ask price levels across limit order books in a couple way).
Finally, closest to our paper is the recent work \cite{MBEB:17} that looks at portfolio execution with a linear cost model with cross-impact terms; their analysis predicates that the portfolio impact matrix has the same eigenvectors as the return correlation matrix, and is stationary. The problem structure allows for their model to be estimable -- in a way similar to what we suggest in our paper, and the stationary model leads to a separable optimal trading schedule, which agrees with our results in that special case.

Our stylized derivation of a price impact model, uses ideas from the market microstructure literature. 
Specifically, as in \cite{Kyle85}, we assume that each investor's holding position on a particular security changes linearly with price, usually justified under a CARA utility function;
the price is determined through a market clearing (equilibrium) condition among all participants.
We don't explicitly specify the trading volume generating process in this paper, the literature of sequential information arrival (\cite{Copeland76}, \cite{Jennings81} and \cite{TauchenPitt83}) provides an insightful connection between trading volume, return volatility and liquidity. 

While we will not estimate an impact model in this paper, we briefly discuss in the last section how one would go about doing so given a set of proprietary portfolio executions.
We refer to \cite{almgren2005direct} for a procedure to estimate an impact model that allows for a linear permanent component and a possibly non-linear instantaneous component without cross-impact effects. 
\cite{HubermanStanzl04} showed using a no-arbitrage argument that the permanent price impact must be a linear function of the quantity traded; see also \cite{Gatheral10} for an extension of that argument to a setting where market impact is transient with a specific decay function.
\cite{RaV:12} make some interesting practical remarks in relation to this market impact model estimation procedure.

The topic of cross-impact has recently started to be explored, specifically in 
\cite{BMEB:16} and \cite{SchL:17}. 
The first paper postulates and estimates a linear propagator impact model  based on the trade sign imbalance vector in each period.
%, i.e., a model where the vectors of differences between buy and sell trades in each time period are linearly combined to estimate the impact in period $t$; 
The second paper explores the implications of the no arbitrage idea of \cite{HubermanStanzl04} to the structure and magnitude of cross-impact. 

%The functional form and decay kernel of the temporary impact term is not as simple to characterize analytically. The simplest assumption treats that decay as being instantaneous; we adapt this working assumption in this paper. 
%Other alternatives typically allow for exponential or power decay functions.The functional form that specifies the magnitude of the temporary cost is itself typically assumed to be linear or sub-linear function of the speed of trading; stylized analytical arguments and statistical evidence suggest a sub-linear functional form.

An important motivation of our work is the gradual shift of assets under management from active to passive and systematic strategies, and their implication to market behavior and the composition and timing of trading flows. 
This topic has been studied in the financial econometrics literature and is summarized in \cite{BenDavid17}. 
In particular, focusing on the topic of liquidity, which is our main concern, this literature has found a causal relationship between ETF or mutual fund ownership and the commonality in the liquidity of the underlying constituents, e.g., see \cite{Karolyi12}, \cite{Koch16}, \cite{Agarwal17}; the motivation of that cross-sectional dependency is attributed to the arbitrage mechanism of ETFs or the correlated trading of mutual funds.\footnote{
The concentration of trading flows towards the end of the trading day has been a popular topic in the financial press; see, e.g., \cite{WSJ2018}.
}

\subsection{Organization of Paper} 
The remainder of the paper is organized as follows. \S \ref{s-empirical} provides some empirical evidence of the cross-sectional variation of intraday liquidity. 
\S \ref{s-model} derives the functional form of a price impact model that incorporates cross-security impact terms that arise from the presence of portfolio (index-fund) liquidity providers.
It subsequently characterizes the expected execution cost given a portfolio trade schedule. 
\S \ref{s-execution} formulates and solves an optimal portfolio execution problem for a risk-neutral investor,
\S \ref{s-illustration} characterizes the performance gains from the optimal schedule over a separable `VWAP-like' execution that is often used in such a risk-neutral setting. 
We conclude in \S \ref{s-practice} with a brief overview of how to estimate price impact from a set of proprietary record of portfolio transactions, and discuss some additional practical considerations.

%\newpage
\section{Preliminary Empirical Observations}
\label{s-empirical}

To motivate our downstream analysis, we provide some empirical evidence regarding the cross-sectional behavior of intraday trading volume, focusing on the level and intraday variation of the pairwise correlations among trading volumes of the S\&P 500 constituent stocks.
We analyzed 459 stocks ($N=459$), indexed by $i$, that had been constituents of S\&P500 throughout the calendar year of 2017. 
Our dataset contains 241 days ($D=241$) indexed by $d$, excluding days that are known to exhibit abnormal trading activity, namely: 
the FOMC/FED announcement days on 02/01, 03/15, 05/03, 06/14, 07/26, 09/20, 11/01, 12/13, and the half trading days on 07/03, 11/24.

We use a Trade-And-Quote (TAQ) database, and extract all trades, excluding those that a) occur before 09:35 or after 16:00; b) opening auction prints or closing auction prints (COND field contains `O', `Q, `M', or `6'); and c)  trades corrected later (CORR field is not 0, or COND field contains `G' or `Z').
We divide a day into five-minute intervals ($T=77$, 09:35-09:40, $\cdots$ 15:55-16:00) indexed by $t$. 
%The dollar amount traded during a certain time interval can be measured by aggregating the individual transactions. 
We denote by $\text{DVol}_{idt}$ the aggregate notional (\$) volume traded on stock $i$ across all transactions that took place in time interval $t$ on day $d$.
We define $\overline{\text{DVol}}_{it}$ to be the yearly average notional volume traded on stock $i$ in time period $t$,
and $\text{AvgVolAlloc}_t$ to be the cross-sectional average \% of daily volume 
traded in period $t$
(``daily volume'' in this definition accounts for all trading activity between 9:35 and 16:00, excluding auction and corrected prints):
\begin{equation} \label{e-vol-alloc-def}
\overline{\text{DVol}}_{it} \triangleq \frac{1}{D} \sum_{d=1}^D \text{DVol}_{idt},
~~\text{VolAlloc}_{it} \triangleq \frac{ \overline{\text{DVol}}_{it}  }{ \sum_{s=1}^T \overline{\text{DVol}}_{is}  }
~~\mbox{and}~~
\text{AvgVolAlloc}_t \triangleq \frac{1}{N} \sum_{i=1}^N \text{VolAlloc}_{it}.
\end{equation}

For each pair of stocks $(i,j)$ we denote by $\text{Correl}_{ijt}$ the pairwise correlation between the respective intraday notional traded volumes across days for each time period $t$. 
As a measure of cross-sectional dependency, we subsequently calculate the average pairwise correlation over all pairs of stocks:
\begin{eqnarray}
\label{e-pair-correl-def}
	\text{Correl}_{ijt} &\triangleq& \frac{ \sum_{d=1}^D (\text{DVol}_{idt} - \overline{\text{DVol}}_{it} )(\text{DVol}_{jdt} - \overline{\text{DVol}}_{jt} ) }{ \sqrt{ \sum_{d=1}^D (\text{DVol}_{idt} - \overline{\text{DVol}}_{it} )^2 \cdot \sum_{d=1}^D(\text{DVol}_{jdt} - \overline{\text{DVol}}_{jt} )^2 } },  \\
	\text{AvgCorrel}_t &\triangleq& \frac{1}{N(N-1)} \sum_{i \ne j} \text{Correl}_{ijt}.
\end{eqnarray}

\begin{figure}[h]
\centering
\begin{subfigure}{.5\textwidth}
  \centering
  \includegraphics[width=.9\linewidth]{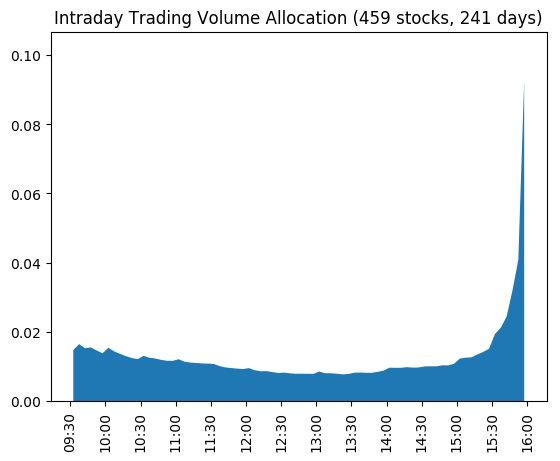}  
\end{subfigure}%
\begin{subfigure}{.5\textwidth}
  \centering
  \includegraphics[width=.9\linewidth]{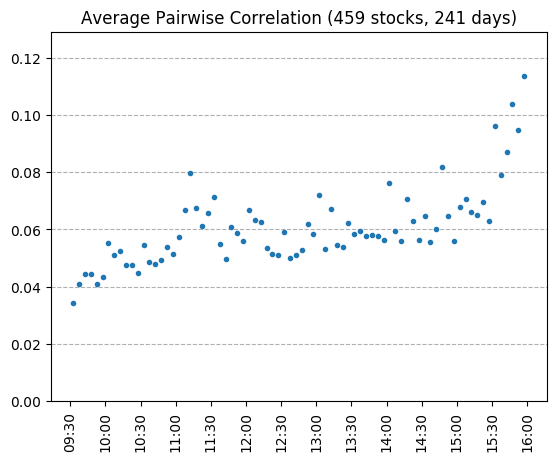}
\end{subfigure}
\caption{Cross-sectional average intraday traded volume profile (left) and cross-sectional average pairwise correlation (right): S\&P 500 constituent stocks in 2017.}
\label{f-empirical}
\end{figure}

Figure \ref{f-empirical} depicts the graphs of $\text{AvgVolAlloc}_t$ and $\text{AvgCorrel}_t$. 
$\text{AvgVolAlloc}_t$ exhibits the commonly observed U-shaped behavior that shows that trading activity is concentrated in the morning and the end of the day. 
The graph of $\text{AvgCorrel}_t$ reveals that (i) trading volumes are positively correlated throughout a day, and (ii) that the cross-sectional average pairwise correlation increases significantly during the last few hours of day.\footnote{Alternative calculations of the intraday volume and correlation patterns produce similar findings.
For example, one could compute stock specific average traded volume profiles, and  for each day compute the stock-specific normalized volume deviation profiles between the realized and forecasted volume profiles; these could be used for the pairwise correlation analysis. 
Similar findings are obtained when we study stocks clustered by their sector, e.g., among financial, energy, manufacturing, etc., stock sub-universes.}

One possible explanation of the observed intraday volume correlation profile could be the non-stationary participation of portfolio order flow throughout the course of the trading session.
Market participants that trade portfolio order flow cause correlated stochastic volume deviations across stocks that, in turn, could contribute to the observed pairwise correlation profile. 
Interpreting portfolio order flows as the primary source of cross-sectional dependency in trading volume, $\text{AvgCorrel}_t$ indirectly reflects the intensity  of portfolio order flow within the total market order flow. 
Our empirical observation indicates that (i) portfolio order flow contributes a certain fraction of trading activity throughout the day, which (ii) is increasing towards the end of the day.
%The increasing proportion could be because the authorized participants of ETFs and managers of passive funds are likely to trade more actively at the end of the day. 
In particular, with the increasing popularity of ETFs and passive funds in recent years, people now trade similar portfolios which may incur stronger cross-sectional dependency; \cite{Karolyi12}, \cite{Koch16} and \cite{Agarwal17} provide empirical evidence showing that the commonality in trading volume indeed arises from the trading activity in ETFs or passive funds.
Similarly, transactions to buy or sell shares into mutual funds are settled to the closing prices, and mutual fund companies tend to execute the net inflows or outflows near or at the end of the trading session.

We will return to these findings about $\text{AvgVolAlloc}_t$ and $\text{AvgCorrel}_t$ in \S \ref{s-illustration} to approximate the relative magnitude of the different type of natural liquidity providers  (portfolio vs. single stock investors), and characterize its effect of incorporating this phenomenon on the optimal execution schedule and execution costs.

\section{Model} \label{s-model}

We assume that there are two types of investors -- single-stock and index-fund investors -- who provide natural liquidity in the market. In this section, we derive the cross-sectional market impact model from a stylized assumption on liquidity provision mechanism of these investors.
The term ``single stock'' here refers to discretionary or active investors that are willing to supply liquidity on individual securities.

\subsection{Single-stock Investors and Index-fund Investors}

Single-stock (discretionary) investors are assumed to trade and provide opportunistic liquidity on individual stocks by adjusting their holdings in response to changes in the fundamental price of the stock. 
This change in single stock investor holdings in stock $i$ is assumed to be linear in the change in the market price with a coefficient $\psi_{\idio,i}$. 
They will sell (or buy) $\psi_{\idio,i}$ shares of stock $i$ when its price $p_i$ rises (or drops) by one dollar.

The assumption about a linear supply relationship between holdings and price is often assumed in the market microstructure literature (\cite{TauchenPitt83}, \cite{Kyle85}). 
It is typically justified under the assumption that risk averse investors choose their holdings to maximize their expected utility given their own belief on the future price. With a CARA utility function and normally distributed beliefs, the optimal holding position is proportional to the gap between current price and their own reservation price, with a proportionality coefficient that incorporates their confidence in their belief and their preference regarding uncertainty. 
Our parameter $\psi_{\idio,i}$ can be thought as a sum of the individual investors' sensitivity parameters.

We consider a universe of $N$ stocks, indexed by $i=1, \ldots, N$.
Suppose that the change in the $N$-dimensional price vector is $\Delta \mathbf{p} \in \mathbb{R}^N$.
Let $\mathbf{e}_i$ be the $i^{th}$ unit vector.
Single-stock investors on stock $i$ will experience the price change $\mathbf{e}_i^\top \Delta \mathbf{p}$ and adjust their holding position by $-\psi_{\idio,i} \cdot \mathbf{e}_i^\top \Delta \mathbf{p}$. 
In vector representation, the change in the holding vector of single-stock investors 
$\Delta \mathbf{h}_\idio \in \mathbb{R}^N$ can be written as
\begin{equation} \label{e-holding-idio}
\Delta \mathbf{h}_{\idio} \left( \Delta \mathbf{p} \right) 
= -\sum_{i=1}^N \mathbf{e}_i \cdot \psi_{\idio,i} \cdot \mathbf{e}_i^\top \Delta \mathbf{p} = -\bm{\Psi}_{\idio} \Delta \mathbf{p}
\quad \in \mathbb{R}^N,
\end{equation}
where
%\begin{equation}
$\bm{\Psi}_{\idio} \triangleq \diag\left( \psi_{\idio,1}, \cdots, \psi_{\idio,N} \right) \in \mathbb{R}^{N \times N}.$
$\Delta \mathbf{h}_{\idio} \left( \Delta \mathbf{p} \right)$ can be thought as ``signed''-volume; i.e., it is positive when orders to buy are submitted in the market when the prices drop, and negative when orders to sell are submitted in the market when prices increase. 
%\end{equation}

%We interpret the coefficient $\psi_{\idio,i}$ as the liquidity on stock $i$ provided by single-stock investors. It basically captures the density of supply or demand per unit price.

In contrast to single-stock investors, index-fund investors trade `portfolios' based on some view on the entire market, a  sector, or a particular group of securities such as high-beta stocks. 
This includes many institutional investors, but the individual investors who hold ETFs or join index funds also belong to this group. 
%With the increasing popularity of passive investment, we expect that there are few portfolios being traded in common, which we call `index funds' later on.
%We particularly focus on passively managed portfolios whose constituents are determined in a systemic way whereas an actively managed fund can be seen as a series of single stock investment.
We assume that there are $K$ such funds, indexed by $k=1, \ldots, K$. 
Let $\mathbf{w}_k = (w_{k1},\cdots,w_{kN})^\top \in \mathbb{R}^N$ be the weight vector of index fund $k$, expressed in \# of shares: one unit of index fund $k$ contains $w_{k1}$ shares of stock 1, $w_{k2}$ shares of stock 2, and so on. 
Given a price change $\Delta \mathbf{p} \in \mathbb{R}^N$, investors on index fund $k$ will experience the price change $\mathbf{w}_k^\top \Delta \mathbf{p}$.
%\begin{equation}
%\text{Price change of index fund $k$ : } \mathbf{w}_k^\top \Delta \mathbf{p}
%\end{equation}
Analogous to single-stock investors, index-fund investors adjust their holding position on index fund $k$ linearly to its price change $\mathbf{w}_k^\top \Delta \mathbf{p}$ with a coefficient $\psi_{\fund,k}$. 
Since trading one unit of index fund $k$ is equivalent to trading a basket of individual stocks with weight vector $\mathbf{w}_k$, we can state the change in index-fund investors' holding position vector  $\Delta \mathbf{h}_\fund \in \mathbb{R}^N$ 
as a vector of changes in the constituents of that fund:
\begin{equation} \label{e-holding-fund}
\Delta \mathbf{h}_{\fund} \left( \Delta \mathbf{p} \right) 
= -\sum_{k=1}^K \mathbf{w}_k \cdot \psi_{\fund,k} \cdot \mathbf{w}_k^\top \Delta \mathbf{p}
= -\W \bm{\Psi}_{\fund} \W^\top \Delta \mathbf{p}
\quad \in \mathbb{R}^N,
\end{equation}
where
\begin{equation}
\W \triangleq \left[ \begin{array}{ccc} \vert & & \vert \\ \mathbf{w}_1 & \cdots & \mathbf{w}_K \\ \vert & & \vert \end{array} \right] \in \mathbb{R}^{N \times K}
, \quad
\bm{\Psi}_{\fund} \triangleq \diag\left( \psi_{\fund,1}, \cdots, \psi_{\fund,K} \right) \in \mathbb{R}^{K \times K}.
\end{equation}

%Combining both types of investors, the price change $\Delta \mathbf{p}$ leads to the change in holding positions in total as follow:
%	\begin{equation}
%		\Delta \mathbf{h}_{\idio} \left( \Delta \mathbf{p} \right) + \Delta \mathbf{h}_{\fund} \left( \Delta \mathbf{p} \right)
%			= - \left( \bm{\Psi}_{\idio} + \W \bm{\Psi}_{\fund} \W^\top \right) \Delta \mathbf{p}
%	\end{equation}

Throughout the paper, we assume that all $\psi_{\idio,i}$'s and $\psi_{\fund,k}$'s are strictly positive, and that $\mathbf{w}_k$'s are linearly independent.

\subsection{Cross-sectional Price Impact} \label{s-price-impact}

Consider a time period over which we wish to execute $\mathbf{v} \in \mathbb{R}^N$ shares. 
Each component can be positive or negative depending on whether we want to buy or sell. 
Our orders transact (eventually) against natural liquidity providers provided by single-stock and index-fund investors;
market makers or high frequency traders tend to maintain negligible inventories at the end of the day, so irrespective of who intermediates the market, we need to elicit $\mathbf{v}$ shares from those two groups of investors. 
A price change of $\Delta \mathbf{p} \in \mathbb{R}^N$ will affect an inventory change of $v$ shares if the following market clearing condition is satisfied:
\begin{equation}
\mathbf{v} + \Delta \mathbf{h}_{\idio} \left( \Delta \mathbf{p} \right) + \Delta \mathbf{h}_{\fund} \left( \Delta \mathbf{p} \right) = \mathbf{0}.
\end{equation}
Based on equations \eqref{e-holding-idio} and \eqref{e-holding-fund},
\begin{equation}
\mathbf{v} 
= \left( \sum_{i=1}^N \mathbf{e}_i \cdot \psi_{\idio,i} \cdot \mathbf{e}_i^\top + \sum_{k=1}^K \mathbf{w}_k \cdot \psi_{\fund,k} \cdot \mathbf{w}_k^\top \right) \Delta \mathbf{p}
= \left( \bm{\Psi}_{\idio} + \W \bm{\Psi}_{\fund} \W^\top \right) \Delta \mathbf{p}.
\end{equation}
The expression means that, out of $\mathbf{v}$ shares, $\bm{\Psi}_\idio \Delta \mathbf{p} \in \mathbb{R}^N$ shares are obtained from single-stock investors and $\W \bm{\Psi}_{\fund} \W^\top \Delta \mathbf{p} \in \mathbb{R}^N$ shares from index-fund investors. This linear relationship between $\mathbf{v}$ and $\Delta \mathbf{p}$ can be translated into the price impact summarized in the next proposition.

\begin{proposition}[Cross-sectional price impact] \label{p-price-impact} 
When executing $\mathbf{v} \in \mathbb{R}^N$ shares, 
the market clearing price change vector $\Delta \mathbf{p} \in \mathbb{R}^N$ is such that 
\begin{equation} \label{e-price-impact}
\Delta \mathbf{p} = \mathbf{G} \mathbf{v}
\quad \text{and} \quad
\mathbf{G} \triangleq \left( \bm{\Psi}_{\idio} + \W \bm{\Psi}_{\fund} \W^\top \right)^{-1}.
\end{equation}
\end{proposition}

Note that the coefficient matrix $\mathbf{G}$ is an inverse of $\bm{\Psi}_{\idio} + \W \bm{\Psi}_{\fund} \W^\top$ which is composed of two symmetric and strictly positive-definite matrices.
Therefore, $\mathbf{G}$ is itself a well-defined symmetric positive-definite matrix, with the following structure: a diagonal matrix plus a non-diagonal low-rank matrix. The following matrix expansion derived from an application of Woodbury's identity will prove useful:
\begin{equation}
\mathbf{G} 
= \left( \underbrace{ \bm{\Psi}_{\idio} }_\text{diagonal} + \underbrace{ \W \bm{\Psi}_{\fund} \W^\top }_\text{rank $K$} \right)^{-1}
= \underbrace{ \bm{\Psi}_{\idio}^{-1} }_\text{diagonal} - \underbrace{ \bm{\Psi}_{\idio}^{-1} \W \left( \bm{\Psi}_{\fund}^{-1} + \W^\top \bm{\Psi}_{\idio}^{-1} \W \right)^{-1} \W^\top \bm{\Psi}_{\idio}^{-1} }_\text{rank $K$} .
\end{equation}

Proposition \ref{p-price-impact} characterizes the structure of the cross price impact model.
The cross-impact is captured by the non-diagonal entries in
$\W \bm{\Psi}_{\fund} \W^\top$ that result as a consequence of the natural liquidity provision attributed to index-fund (portfolio) investors. 
%Apart from mathematical formulation, it naturally comes out from the fact that a certain fraction of market participants are who trade portfolios. 
%Even if we trade a single stock, it will trigger trades in the other securities as long as portfolio traders participate as a counter party, and hence their price will move consequently.

%\textbf{(Interpretation of liquidity variable $\psi$)}
We shall interpret the terms $\bm{\Psi}_{\idio} \triangleq \text{diag}_{i=1}^N( \psi_{\idio,i} )$ and $\bm{\Psi}_{\fund} \triangleq \text{diag}_{k=1}^K( \psi_{\fund,k} )$ as \textit{``liquidity''}. $\psi_{\idio,i}$ represents the 
amount of liquidity provided by single stock investors in stock $i$ and $\psi_{\fund,k}$ represents the liquidity supplied by of index fund $k$ investors. 
The sum $\bm{\Psi}_{\idio} + \W \bm{\Psi}_{\fund} \W^\top$ indicates the total market liquidity.
As shown in \eqref{e-price-impact}, the price impact is inversely proportional to the liquidity, which agrees with the conventional definition of liquidity as a measure of ease of trading. 
When $\psi_{\idio,i}$ or $\psi_{\fund,k}$ is large, equivalently when the liquidity is abundant, price impact is low. 
%Such a relationship can be understood from its definition. 
$\psi_{\idio,i}$ and $\psi_{\fund,k}$ were originally defined as the sensitivity of investors' holdings to market price movements, and, as such,
capture how many shares we can obtain from these two types of investors when the price moves by a certain amount; a measure of price impact.

\iffalse
The value of $\psi_{\idio,i}$ (or $\psi_{\fund,k}$) could be decomposed further: (i) the number of investors who participate while we are executing, and (ii) the sensitivity of each individual investor's holding position to market price. An abundant liquidity could arise because many participants are actively trading or because each investor provides a lot of liquidity. The first component reflects the \textit{trading activity} of market participants, which is also closely related to \textit{trading volume}. And the second component reflects some behavioral characteristics of investor, which might be theorized with market microstructure theory. Even though we don't formulate this decomposition explicitly, we shall rely on this understanding throughout the paper. For example, we use trading volume as a proxy for the first component in Section \ref{example} and Section \ref{practice}, and we relate volatility with the second component in Section \ref{practice}.
\fi

\subsection{One-period Transaction Cost}

Consider a liquidator that wishes to execute $\mathbf{v} \in \mathbb{R}^N$ shares over a short period of time, say 5 to 15 minutes.
Let $\mathbf{p}_0 \in \mathbb{R}^N$ be the price at the beginning of this execution period.
Assuming that $\mathbf{v}$ is traded continuously and at a constant rate over the duration of that time period, the liquidator will realize an average transaction price given by
\begin{equation} \label{e-avg-tr-prc}
\mathbf{\bar{p}}^\tran 
= \mathbf{p}_0 + \frac{1}{2} \mathbf{G} \mathbf{v} + \bar{\bm{\epsilon}}^\tran ,
\end{equation}
where $\bar{\bm{\epsilon}}^\tran \in \mathbb{R}^N$ represents a random error term that captures unpredictable market price fluctuations or the effect of trades executed in that period by other investors;
this suggests that costs accumulate linearly over the duration of the period, and that the average execution price is halfway the average impact plus a random contribution due to fluctuations in the price due to exogenous factors. (We will return to this assumption later on.)
We will assume that the error is independent of our execution $\mathbf{v}$ and zero mean: i.e., $\E\left[ \bar{\bm{\epsilon}}^\tran | \mathbf{v} \right] = \mathbf{0}$.
The single-period expected implementation shortfall incurred by the liquidator 
is given by
\begin{equation}
\bar{\cal C}\left( \mathbf{v} \right)
\triangleq \E\left[ \mathbf{v}^\top \left( \mathbf{\bar{p}}^\tran - \mathbf{p}_0 \right) \right] = \frac{1}{2} \mathbf{v}^\top \mathbf{G} \mathbf{v}.
\end{equation}
Linear price impact induces quadratic implementation shortfall costs; note that the resulting cost is always positive since $\mathbf{G}$ is positive definite. 
The following proposition briefly explores how the mixture of natural liquidity providers affects the expected execution cost. 

\begin{proposition}[Extreme cases] \label{p-tr-cost}
Consider a parametric scaling of the single-stock and index-fund natural liquidity, 
 $\bm{\Psi}_\idio$ and $\bm{\Psi}_\fund$, respectively, given by 
 \begin{equation}
\mathbf{G} = \left( \alpha \cdot \bm{\Psi}_\idio + \beta \cdot \W \bm{\Psi}_\fund \W^\top \right)^{-1},
\end{equation}
for some scalars $\alpha \in (0,1]$ and $\beta \in (0,1]$.

(i) If there are no index-fund investors ($\alpha \rightarrow 1$ and $\beta \rightarrow 0$), the expected execution cost becomes separable across individual assets:
\begin{equation} \label{e-tr-cost-idio}
\lim_{\alpha \rightarrow 1, \beta \rightarrow 0} \bar{\cal C}\left( \mathbf{v} \right) 
= \frac{1}{2} \mathbf{v}^\top \bm{\Psi}_\idio^{-1} \mathbf{v}
= \frac{1}{2} \sum_{i=1}^N \frac{ v_{i}^2 }{ \psi_{\idio,i} }.
\end{equation}

(ii) If there are no single-stock investors ($\alpha \rightarrow 0$ and $\beta \rightarrow 1$), the liquidator can only execute with finite expected execution costs portfolio orders that can be expressed as a linear combination of the index-fund weight vectors. Specifically:
\begin{equation} \label{e-tr-cost-fact}
\lim_{\alpha \rightarrow 0, \beta \rightarrow 1} \bar{\cal C}\left( \mathbf{v} \right) 
= \left\{ \begin{array}{cl}
\infty	 & \textup{if } \mathbf{v} \notin \textup{span}\left( \mathbf{w}_1, \cdots, \mathbf{w}_K \right) \\
\frac{1}{2} \mathbf{u}^\top \bm{\Psi}_{\fund}^{-1} \mathbf{u} 	& \textup{if } \mathbf{v} = \mathbf{W} \mathbf{u}
\end{array} \right. .
\end{equation}
\end{proposition}

(The proof is provided in Appendix \ref{prf-tr-cost}.)
Therefore, separable (security-by-security) market impact cost models, often assumed in practice, essentially predicate, per our analysis, that all natural liquidity in the market is provided by opportunistic single-stock investors.
And, in that case, \eqref{e-tr-cost-idio} recovers the commonly used ``diagonal'' market impact cost model.
The other extreme scenario assumes that all liquidity is provided along the weight vectors of the index fund investors, and the resulting cost then depends on how the target execution vector $\mathbf{v}$ can be expressed as a linear combination of the $(\mathbf{w}_1, \cdots, \mathbf{w}_K )$.
In practice, the latter case suggests that execution costs may increase in periods with relatively higher intensity of portfolio liquidity provision when the target portfolio that is being liquidated is not well aligned with the directions in which portfolio liquidity is supplied.

\subsection{Time-varying Liquidity and Multi-period Transaction Cost}

The stylized observations of Proposition \ref{p-tr-cost} suggest that intraday trading costs may be affected by intraday variations in the mixture of natural liquidity providers, and, in particular, if the relative contribution of index fund investors increases significantly over time. 

We will consider the transaction cost of an intraday execution schedule $\mathbf{v}_1, \cdots, \mathbf{v}_T$ over $T$ periods, in which $\mathbf{v}_t \in \mathbb{R}^N$ shares are executed during the time interval $t$. 
We will make the following assumptions on the intraday behavior of price impact, price dynamics, and realized execution costs.

a)  We allow the mixture of liquidity provision to fluctuate over the course of the day. We denote the time-varying liquidity by $\psi_{\idio,it}$ and $\psi_{\fund,kt}$ with an additional subscript $t$. 
We assume that the portfolio weight vectors $\mathbf{w}_k$ of index liquidity providers are assumed to be fixed during a given day. 
Under this setting, the coefficient matrix of price impact can be represented as follow:
\[
\mathbf{G}_t = \left( \bm{\Psi}_{\idio,t} + \W \bm{\Psi}_{\fund,t} \W^\top \right)^{-1}. 
\]
%We attribute the intraday variation of liquidity to the variation in number of investors who are actively participating the market.

b) Let $\mathbf{p}_t$ be the fundamental price at the end of period $t$. The `fundamental' price means the price on which the market agrees as a best guess of the future price excluding the temporary deviation of the realized transaction price due to market impact. 
The fundamental price process $(\mathbf{p}_0, \mathbf{p}_1, \cdots, \mathbf{p}_T)$ is assumed to be a martingale independent from the execution schedule:
\[
\mathbf{p}_t = \mathbf{p}_{t-1} + \bm{\epsilon}_t, \quad \text{for all } t = 1, \cdots, T,
\]
where the innovation term $\bm{\epsilon}_t$ satisfies $\E\left[ \bm{\epsilon}_t | {\cal F}_{t-1} \right] = \mathbf{0}$ with all the past information ${\cal F}_{t-1}$. The term $\bm{\epsilon}_t$ is commonly understood as the change in market participants' belief perhaps due to the information revealed during the period $t$. We are implicitly assuming that our execution conveys no information about the future price.
		
c) The realized `transaction' price in each period can deviate from the fundamental price temporarily, e.g., due to a short-term imbalance between buying order flow and selling order flow. 
In executing $\mathbf{v}_t$ shares, the liquidator is contributing to such an imbalance, which causes the temporary price impact according to the mechanism described earlier on. 
We assume that this impact is temporary, we particularly assume that the transaction price begins at the fundamental price at each period regardless of the liquidator's trading activity in prior periods. 
Given the coefficient matrix $\mathbf{G}_t$, when $\mathbf{v}_t$ is executed smoothly, the average transaction price is
\[
\mathbf{\bar{p}}^\tran_t = \mathbf{p}_{t-1} + \frac{1}{2} \mathbf{G}_t \mathbf{v}_t + \bar{\bm{\epsilon}}^\tran_t,
\]
where the error term $\bar{\bm{\epsilon}}^\tran_t$ satisfies $\E\left[ \bar{\bm{\epsilon}}^\tran_t | \mathbf{v}_t \right] = \mathbf{0}$ as before.

Under these assumptions, the expected transaction cost of executing a series of portfolio transactions $\mathbf{v}_1, \cdots, \mathbf{v}_T$ is separable over time and can be expressed as follows: 
\[
\bar{\cal C}\left( \mathbf{v}_1, \cdots, \mathbf{v}_T \right)
\triangleq \E\left[ \sum_{t=1}^T \mathbf{v}_t^\top \left( \mathbf{\bar{p}}^\tran_t - \mathbf{p}_0 \right) \right]
= \sum_{t=1}^T \frac{1}{2} \mathbf{v}_t^\top \mathbf{G}_t \mathbf{v}_t.
\]
This formulation implicitly assumes that the intraday liquidity captured through 
$\psi_{\idio,it}$'s and $\psi_{\fund,kt}$'s is deterministic and known in advance. 
Although intraday liquidity evolves stochastically over the course of the day, its expected profile exhibits a fairly pronounced shape that serves as a forecast that can be used as a basis for analysis (as is done in practice); c.f., the discussion in \S \ref{s-practice}.

\section{Optimal Portfolio Execution}	\label{s-execution}

We will formulate and solve the multi-period optimal portfolio execution problem in \S \ref{ss-opt-sol}, and subsequently explore the properties of the optimal solution as a function of the intraday variation of the two sources of natural liquidity providers in \S \ref{ss-structure}.

\subsection{Optimal Trade Schedule} \label{ss-opt-sol}

Consider a risk neutral liquidator interested in executing $\mathbf{x}_0 \in \mathbb{R}^N$ shares over an execution horizon $T$ (e.g., a day). 
We formulate a discrete-time optimization problem to find an optimal schedule $\mathbf{v}_1, \cdots, \mathbf{v}_T$ that minimizes the expected total transaction cost:
\begin{eqnarray} \label{e-opt-prob}
\text{minimize} && \bar{\cal C}\left( \mathbf{v}_1, \cdots, \mathbf{v}_T \right) = \sum_{t=1}^T \frac{1}{2} \mathbf{v}_t^\top \mathbf{G}_t \mathbf{v}_t \\
\text{subject to} && \sum_{t=1}^T \mathbf{v}_t = \mathbf{x}_0.
\label{e-opt-prob2}
\end{eqnarray}
% The above formulation implicitly assumes that the liquidity variation over execution horizon is known in advance. If someone assumes stochastic evolution of liquidity, predicted values can be used. 

\begin{proposition}[``Coupled'' execution] \label{p-opt-sol} The risk-neutral cost minimization problem \eqref{e-opt-prob}-\eqref{e-opt-prob2} has a unique optimal solution given by
\begin{equation} \label{e-opt-sol}
\mathbf{v}_t^* 
~=~ \mathbf{G}_t^{-1} \left( \sum_{s=1}^T \mathbf{G}_s^{-1} \right)^{-1} \mathbf{x}_0 ~=~ \left( \bm{\Psi}_{\idio,t} + \W \bm{\Psi}_{\fund,t} \W^\top \right) 
\left( \Pid + \W \Pf \W^\top \right)^{-1} \mathbf{x}_0,
\end{equation}
where the total daily liquidity $\Pid$ and $\Pf$ are defined as follows
\begin{equation}
\Pid \triangleq \sum_{t=1}^T \bm{\Psi}_{\idio,t}
, \quad
\Pf \triangleq \sum_{t=1}^T \bm{\Psi}_{\fund,t}.
\end{equation}
\end{proposition}

We make the following observations.
First, the optimal solution is ``coupled'' across securities. 
Specifically, as long as the market impact is cross-sectional, the cost minimizing solution needs to consider all orders simultaneously in optimally scheduling how to liquidate the constituent orders, as opposed to scheduling each order separately and attempting to minimize costs as if market impact is separable;
such a separable execution approach is often used in practice (effectively assuming that there are no cross impact effects).
The coupled execution recognizes that the blend of natural liquidity changes intraday, and attempts to change the composition of the residual liquidation portfolio so as to take advantage of portfolio liquidity that may be available towards the end of the day, for example.
We will explore this point further in the remainder of this section.
Second, it is typical to derive coupled optimal portfolio trade schedules for investors that are risk-averse and consider the variance of the execution costs in the objective function or as a constraint; in that case, the covariance structure of the portfolio throughout its liquidation horizon intuitively leads to coupled execution solution.
In our problem formulation, the coupling of the execution path is driven by the cross-sectional dependency of natural (portfolio) liquidity provided by index funds that leads to cross-impact, as opposed to the cross-sectional dependency of intraday returns. 
Third, we note that in the above formulation we have not imposed side constraints that would enforce that the liquidation path is monotone; we will return to this point later on.

The structure of the optimal schedule in \eqref{e-opt-sol} takes an intuitive form:
the proportion of the trade that is liquidated in period $t$ is proportional to the available liquidity in that period, as captured by the time-dependent numerator matrix $\bm{\Psi}_{\idio,t} + \W \bm{\Psi}_{\fund,t} \W^\top$, normalized by the total  liquidity made available throughout the day, as captured by the time-independent denominator matrix $\Pid + \W \Pf \W^\top$.
An alternative interpretation also given by \eqref{e-opt-sol} is that the optimal schedule splits the order inversely proportional to a normalized time-dependent market impact matrix.

\begin{corollary}[No index fund investors, $\bm{\Psi}_{\fund,t} =0$ for $t=1, \ldots, T$] 
\label{coro-no-index-funds}
When there are no index fund investors: i.e., $\bm{\bar{\psi}}_\fund = \mathbf{0}$, a separable `VWAP'-like trade schedule is optimal:
\begin{equation}
v_{it}^* = \frac{ \psi_{\idio,it} }{ \sum_{s=1}^T \psi_{\idio,is} } \cdot x_{i0}, \quad
\text{for } i = 1, \cdots, N.
\end{equation}
\end{corollary}

\noindent
\proofnamest{Proof of Proposition \ref{p-opt-sol}}
Note that since $\mathbf{G}_t$ is symmetric, $\frac{\partial}{\partial \mathbf{v}_t} \frac{1}{2} \mathbf{v}_t^\top \mathbf{G}_t \mathbf{v}_t = \mathbf{G}_t \mathbf{v}_t$.
The KKT conditions of the convex minimization problem in 
\eqref{e-opt-prob}-\eqref{e-opt-prob2}  
require that there exists a vector $\bm{\lambda} \in \mathbb{R}^N$ such that
\[
\bm{\lambda} 
= \left. \frac{\partial}{\partial \mathbf{v}_t} \frac{1}{2} \mathbf{v}_t^\top \mathbf{G}_t \mathbf{v}_t \right|_{\mathbf{v}_t = \mathbf{v}_t^*} 
= \mathbf{G}_t \mathbf{v}_t^*, \quad \text{for all } t = 1, \cdots, T,
\]
which together with the inventory constraint in \eqref{e-opt-prob2} imply that
\[
\mathbf{x}_0 = \sum_{t=1}^T \mathbf{v}_t^* = \sum_{t=1}^T \mathbf{G}_t^{-1} \bm{\lambda}.
\]
It follows that $\mathbf{v}_t^* = \mathbf{G}_t^{-1} \bm{\lambda} = \mathbf{G}_t^{-1} \left(\sum_{s=1}^T \mathbf{G}_s^{-1} \right)^{-1} \mathbf{x}_0$. Since all $\mathbf{G}_t$'s are invertible, the optimal solution exists and is unique. \qed

In a market where all natural liquidity is provided by single stock, opportunistic investors, there are not cross-security impact effects, market impact is separable, and the minimum cost schedule for a risk-neutral liquidator is also separable across securities -- the optimal solution simply needs to minimize expected impact costs separately for each order in the portfolio. 
Each individual order can be scheduled independently of the others, and the resulting schedule is `VWAP'-like in that the execution quantity $v_{it}$ is proportional to the available liquidity $\psi_{\idio,it}$ at that moment. 
Indeed, treating the overall market trading volume profile as the observable proxy of the natural liquidity profile, the solution spreads each order separately and in a way that is proportional to the percentage of the market volume that is forecasted for  each time period; this is what a typical VWAP execution algorithm does. 

%So, similarly to our earlier observations, if there are no index fund (portfolio) liquidity providers, market impact costs are separable and can be optimized on a security-by-security basis, and a separable (VWAP) schedule is optimal.
Conversely, if some of the natural liquidity is provided by index fund investors that wish to trade portfolios --e.g., liquidate some amount of an energy tracking portfolio if the energy sector has had a significant, positive return intraday, we would expect that the separable VWAP schedule does not minimize expected market impact costs, and it is not optimal for the motivating trade scheduling problem.

\subsection{Optimal Trade Schedule under Parametric Liquidity Profile}
\label{ss-structure}

%We've derived the optimal execution schedule under a very general specification in which liquidity can vary arbitrarily. To deepen our understanding about the optimal schedule, we impose some hypothetical structure on intraday variation of liquidity and characterize its implications through the theoretical analysis.

To gain some insight on the structure of the optimal policy we explore a setting where the intensity of single stock and index fund liquidity provision varies parametrically as follows:
single-stock investors' liquidity $\psi_{\idio,it}$ varies over time $t=1,2,\cdots,T$ according to a profile $\alpha_t$, and index-fund investors' liquidity $ \psi_{\fund,kt} $ varies according to another profile $\beta_t$.
\begin{equation} \label{e-alpha-beta}
\bm{\Psi}_{\idio,t} = \alpha_t \cdot \bm{\bar{\Psi}}_\idio, \quad \bm{\Psi}_{\fund,t} = \beta_t \cdot \bm{\bar{\Psi}}_\fund, \quad \text{for } t = 1, \cdots, T,
\end{equation}
where $\sum_{t=1}^T \alpha_t = \sum_{t=1}^T \beta_t = 1$. 

We will assume that all single stocks share the same time-varying profile $\alpha_t$, and likewise all index funds share the profile $\beta_t$.
%the two profiles are allowed to be different.
The empirical findings of \S \ref{s-empirical} indicate that pairwise correlations of trading volumes increase towards the end of the day. 
If a primary source of stochastic fluctuations in intraday trading volumes is the stochastic arrivals of single stock and portfolio trades, then one would expect that the profiles $\alpha_t, \beta_t$ vary intraday so as to generate the well known $U$-shaped volume profile, and to vary differently from each other so as as to generate the time-varying pairwise correlation relationship; this is supported by behavior of market participants towards the end-of-day, discussed earlier.
Indeed, if the two groups of natural liquidity had the same trading activity profiles, i.e., $\alpha_t = \beta_t$, then the average correlation in intraday trading volume would not vary intraday. 
We expect that towards the end of the day, the intensity of index fund liquidity provision ($\beta_t$) increases relatively faster than the intensity of single stock liquidity provision ($\alpha_t$).
%at the end of the day as the correlation increases. See Section \ref{example} for the actual values of $\alpha_t$'s and $\beta_t$'s retrieved from the empirical observation. 

\begin{proposition}[Optimal execution under structured variation] \label{p-tilting}
Under the parameterization of \eqref{e-alpha-beta}, the schedule $\mathbf{v}_t^*$ is optimal for the risk-neutral cost minimization \eqref{e-opt-prob}:
\begin{equation}
\mathbf{v}_t^* = \alpha_t \cdot \mathbf{x}_0 + (\beta_t - \alpha_t) \cdot \W \left( \Pf^{-1} + \W^\top \Pid^{-1} \W \right)^{-1} \W^\top \Pid^{-1} \mathbf{x}_0,
\end{equation}
or, equivalently, 
\begin{equation} \label{e-tilting}
\mathbf{v}_t^* = \alpha_t \cdot \mathbf{x}_0 + (\beta_t - \alpha_t) \cdot \sum_{k=1}^K (\mathbf{\widehat{w}}_k^\top \mathbf{x}_0) \cdot \mathbf{w}_k,
\end{equation}
where $ \mathbf{\widehat{W}} \triangleq \Pid^{-1} \W \left( \Pf^{-1} + \W^\top \Pid^{-1} \W \right)^{-1} $,
and $\mathbf{\widehat{w}}_k$ denotes the $k^{th}$ column of 
$\mathbf{\widehat{W}}$.
\end{proposition}

Before offering an interpretation for \eqref{e-tilting} we state the following corollary. 

\begin{corollary}[Optimal execution under common variation] \label{cor-common-variation}
If $\alpha_t = \beta_t$ for all $t=1,\cdots,T$, a separable, `VWAP'-like strategy is again optimal.
\begin{equation} \label{e-common-variation}
\mathbf{v}_t^* = \alpha_t \cdot \mathbf{x}_0.
\end{equation}
\end{corollary}

The proof of Proposition \ref{p-tilting} is given in the Appendix \ref{prf-tilting}. Corollary \ref{cor-common-variation} states that when the intensity of natural liquidity provision is the same for single stock and index fund investors, $\alpha_t = \beta_t$, the optimal schedule $\mathbf{v}_t^*$ is again aligned with $\mathbf{x}_0$ scaled by 
$\alpha_t$. 
As $\alpha_t$ ($=\beta_t$) represents the market activity at time $t$, the above policy can be interpreted as a `VWAP'-like execution that spreads each individual orders proportionally to the total volume available at each point in time; this is separable across orders.  
As noted earlier, the setting where $\alpha_t = \beta_t$ is, however, inconsistent with the empirical findings regarding the intraday behavior of pairwise correlations of trading volumes.
%When $\alpha_t = \beta_t$ so the cross-sectional dependency in intraday liquidity doesn't exhibit any intraday variation, the optimal schedule results in a trivial solution.

In contrast, \eqref{e-tilting} highlights that when the mixture of natural liquidity varies intraday (through the difference between $\alpha_t$ and $\beta_t$) , the optimal schedule tilts away from the `VWAP'-like execution encountered in \eqref{e-common-variation} so as to take advantage of increased available index fund liquidity, e.g., offered along the direction of sector portfolios.
%As an example, consider the liquidation of a portfolio of technology positions and suppose that there is increased natural liquidity provided towards the end of the trading session along the direction of the sector portfolio; i.e., trading along the sector weights is relatively cheaper towards the end of the day. If the liquidation portfolio is significantly different than the technology sector portfolio, the optimal policy is to trade early in the day in a way that the residual portfolio towards the end of the day is more closely aligned with the sector portfolio that is cheaper to trade.
%changes intraday, It takes the `naive' VWAP execution $\alpha_t \cdot \mathbf{x}_0$ as a default and it tilts toward (or tilts away) the direction of index funds $\mathbf{w}_k$ when $\beta_t > \alpha_t$ (or $\beta_t < \alpha_t$). The amount of tilting also depends on how much $\mathbf{x}_0$ is aligned with the index funds. In words, it's better to execute along the direction of index funds when index-fund investors are relatively more active than single-stock investors.

\section{Illustration of Optimal Execution and Performance Bounds}
\label{s-illustration}

In this section we provide a brief illustration of the optimized execution path that incorporates the effect of index fund (portfolio) liquidity.
Risk-neutral investors often adapt a separable execution style, i.e., trade each asset separately, most often using a Volume-Weighted-Average-Price (VWAP) algorithm.
As we show in \S \ref{s-execution}, this separable strategy, under some assumptions, can be shown to minimize expected impact costs per order, but disregards the effect of portfolio liquidity and cross-impact costs when trading multiple orders side-by-side.
For a stylized model of natural liquidity of the form introduced in  \S \ref{ss-structure} simplified to the case of a single index fund (e.g., the market portfolio), 
we establish a worst case bound on the sub-optimality gap of such a separable execution approach against the optimized portfolio schedule derived earlier.

Specifically, restricting attention to the parameterization introduced in \S \ref{ss-structure} in a setting with a single index fund ($K=1$):
$\bm{\Psi}_{\idio,t} = \alpha_t \cdot \Pid$ and $\bm{\Psi}_{\fund,t} = \beta_t \cdot \Pf$ with $\sum_{t=1}^T \alpha_t = \sum_{t=1}^T \beta_t = 1$, and
Proposition \ref{p-tilting} states that the optimal execution $\mathbf{v}_t^*$ is
\begin{equation} \label{e-opt-v}
\mathbf{v}_t^* = \alpha_t \cdot \mathbf{x}_0 + (\beta_t - \alpha_t) \cdot \left( \mathbf{\widehat{w}}_1^\top \mathbf{x}_0 \right) \cdot \mathbf{w}_1
, \quad \text{for } t =1,\cdots,T ,
\end{equation}
where $\mathbf{w}_1 \in \mathbb{R}^N$ is the weight vector of the index fund (e.g., the market portfolio), expressed in number of shares, and $\mathbf{\widehat{w}}_1 \triangleq \left( \bar{\psi}_{\fund,1}^{-1} + \mathbf{w}_1^\top \Pid^{-1} \mathbf{w}_1 \right)^{-1} \Pid^{-1} \mathbf{w}_1$.
In contrast, the separable execution $\mathbf{v}_t^\text{sep}$ liquidates each order in the portfolio independently, allocating quantities to be traded in each period in a way that is proportional to the total traded volume that is forecasted to execute in that period:
\begin{equation} \label{e-sep-v}
v_{it}^\text{sep} = \text{VolAlloc}_{it} \cdot x_{0i}
, \quad \text{for } t=1,\cdots,T, \quad \text{for each } i=1,\cdots,N ,
\end{equation}
where $\text{VolAlloc}_{it}$ is the percentage of the daily volume in security $i$ that trades in period $t$, defined in \eqref{e-vol-alloc-def}.

\S \ref{ss-volume-profile} (and, in more detail, Appendix \S \ref{app-generative}) posits a stylized stochastic process generative model for single stock and portfolio (index fund) investor order flow, that results in a simple parametric structure for the total traded volume profile $\text{VolAlloc}_{it}$ and the resulting pairwise correlation profile (among traded volumes) $\text{Correl}_{ijt}$.
The model's primitive parameters can be estimated so as to be consistent with the $\text{AvgVolAlloc}_t$ and $\text{AvgCorrel}_t$ depicted in \S \ref{s-empirical}.
\S \ref{ss-perf-bounds} provides analytic results on the optimality gap between the separable and the optimal execution schedules, in (\ref{e-sep-v}) and (\ref{e-opt-v}), respectively, which for the parameters estimated in \S \ref{ss-volume-profile} could be as high as 6.2\%.

\subsection{A Useful Parametrization of Intraday Liquidity}
\label{ss-volume-profile}

We will posit a simple generative model of single stock and portfolio (index fund) order flow (driven by two underlying Poisson processes).
This mixture of order flows comprise the total volume for the day, and also generates a certain correlation structure in the traded volumes per period across securities. (We will offer a brief overview in this section, and defer to the Appendix \S \ref{app-generative} for additional detail on this model.)
Let $\theta_i$ denote the fraction of traded volume in a day for stock $i$ that is generated by order flow submitted by index fund investors.
Formally,
\begin{equation}
\theta_i \triangleq 
%\frac{ \sum_{t=1}^T \E\left[ |\tilde{w}_{1i}| \cdot Q_{\fund,dt} \right] }{ \sum_{t=1}^T \E\left[ \text{DVol}_{idt} \right] }
\frac{ |\tilde{w}_{1i}| \cdot \bar{q}_\fund }{ \bar{q}_{\idio,i} + |\tilde{w}_{1i}| \cdot \bar{q}_\fund },
\end{equation}
where $\bar{q}_\fund$ is the notional traded by portfolio investors, $\tilde{w}_{1i}$ is the weight of security $i$ in this index fund (notionally weighted), and 
$\bar{q}_{\idio,i}$ is the notional traded by single stock investors in security $i$.
For simplicity, further assume that $\theta_1 = \theta_2 = \cdots \theta_N = \theta$, i.e., all securities have the same composition of order flow as contributed by single stock and portfolio (index fund) investors.

In such a model, (as explained in \S \ref{app-generative}) the intraday volume and pairwise correlation profiles are given by:
\begin{eqnarray}
\label{e-theta1}
\text{AvgVolAlloc}_t &=& \frac{1}{N} \sum_{i=1}^N \frac{ \E\left[ \text{DVol}_{idt} \right] }{ \sum_{s=1}^T \E\left[ \text{DVol}_{ids} \right] } = \alpha_t \cdot (1-\theta) + \beta_t \cdot \theta ,
\\
\text{AvgCorrel}_t &=& \frac{1}{N(N-1)} \sum_{i \ne j} \text{Correl}_{ijt} = \frac{\beta_t \cdot \theta^2}{\alpha_t \cdot (1-\theta)^2 + \beta_t \cdot \theta^2}.
\label{e-theta2}
\end{eqnarray}
We note that the assumption that $\theta_i = \theta$ for all securities $i$ leads to the conclusion that all securities have the same intraday volume profile, and, perhaps, more importantly that the intraday volume correlation profile $\text{Correl}_{ijt}$ is the same across all pairs of stocks. 
The latter is arguably a fairly strong restriction, and it is only imposed so as to allow for a more tractable closed form performance analysis.
%Both of these implied restrictions, and, in particular, the latter, are fairly strong, but the resulting parsimonious expressions will allow for a fairly direct illustration of the potential magnitude of the phenomenon we studied in the previous two sections even when relying on public (trade) data. 

\begin{figure}[!h]
\centering
\begin{subfigure}{.5\textwidth}
  \centering
  \includegraphics[width=.9\linewidth]{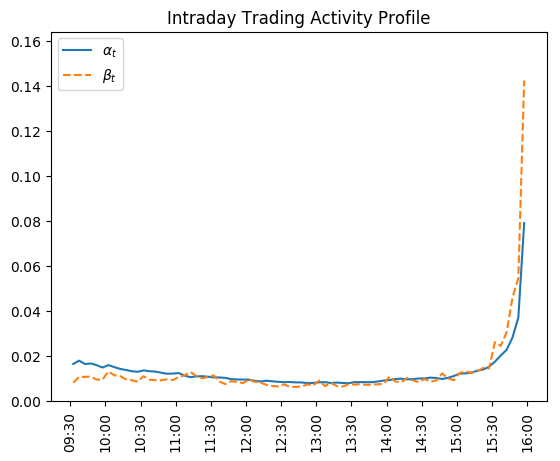}  
\end{subfigure}%
\begin{subfigure}{.5\textwidth}
  \centering
  \includegraphics[width=.9\linewidth]{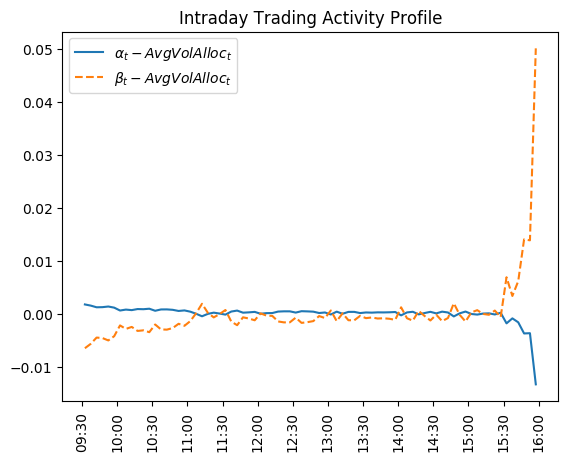}
\end{subfigure}
\caption{Left panel: the intensity of single stock investors, $\alpha_t$, and portfolio (index fund) investors, $\beta_t$, calibrated to best match the empirical profiles of traded volume, $\text{AvgVolAlloc}_t$ and pairwise volume correlations, $\text{AvgCorrel}_t$. Right panel: depicts the deviation of $\alpha_t, \beta_t$ form the market profile $\text{AvgVolAlloc}_t$.}
\label{f-example-profiles}
\end{figure}

\begin{figure}[!h]
\centering
\begin{subfigure}{.5\textwidth}
  \centering
  \includegraphics[width=.9\linewidth]{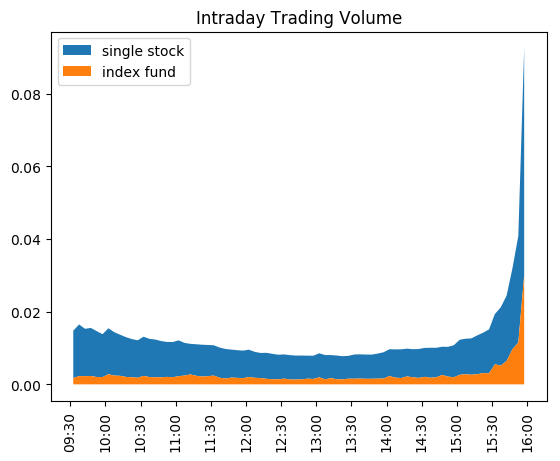}  
\end{subfigure}%
\begin{subfigure}{.5\textwidth}
  \centering
  \includegraphics[width=.9\linewidth]{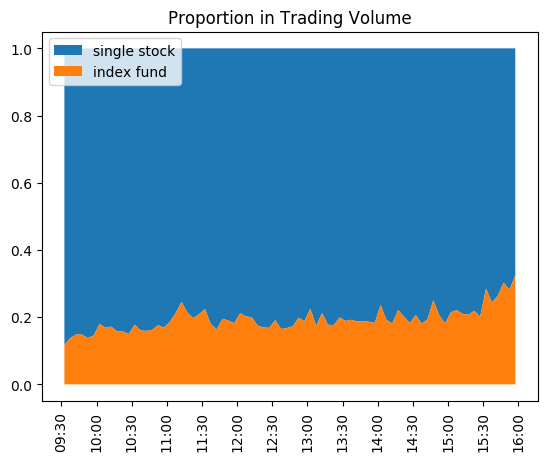}
\end{subfigure}
\caption{The intraday trading volume profiles $\alpha_t \cdot (1-\theta)$ and $\beta_t \cdot \theta$ (left), and the proportion of index-fund order flows $\frac{\beta_t \cdot \theta}{\alpha_t \cdot (1-\theta) + \beta_t \cdot \theta}$ (right). }
\label{f-example-stats}
\end{figure}

Given the empirically observed profiles for $\text{AvgVolAlloc}_t$ and $\text{AvgCorrel}_t$, visualized in Figure \ref{f-empirical}, we can solve a set of coupled equations defined by \eqref{e-theta1}-\eqref{e-theta2} where the respective left hand sides are given by the empirically estimated values,
to identify the values of $\theta$, $\alpha_1, \cdots, \alpha_T$ and $\beta_1, \cdots, \beta_T$.
The results are summarized in Figures \ref{f-example-profiles} and \ref{f-example-stats}.
$\theta$ was estimated to be 0.21, implying that 21\% of total traded volume originate from the index fund.
We can observe that, at the beginning of the day, the trading activity of index-fund investors $\beta_t$ is smaller than that of single-stock investors $\alpha_t$, but $\beta_t$ far exceeds $\alpha_t$ in the last hour of the day, as expected.
Such an intraday variation in the composition of order flow is consistent with the increasing pairwise correlation in volumes towards the end of the trading day. 

\begin{figure}[!h]
	\centering
	\includegraphics[width=0.7\linewidth]{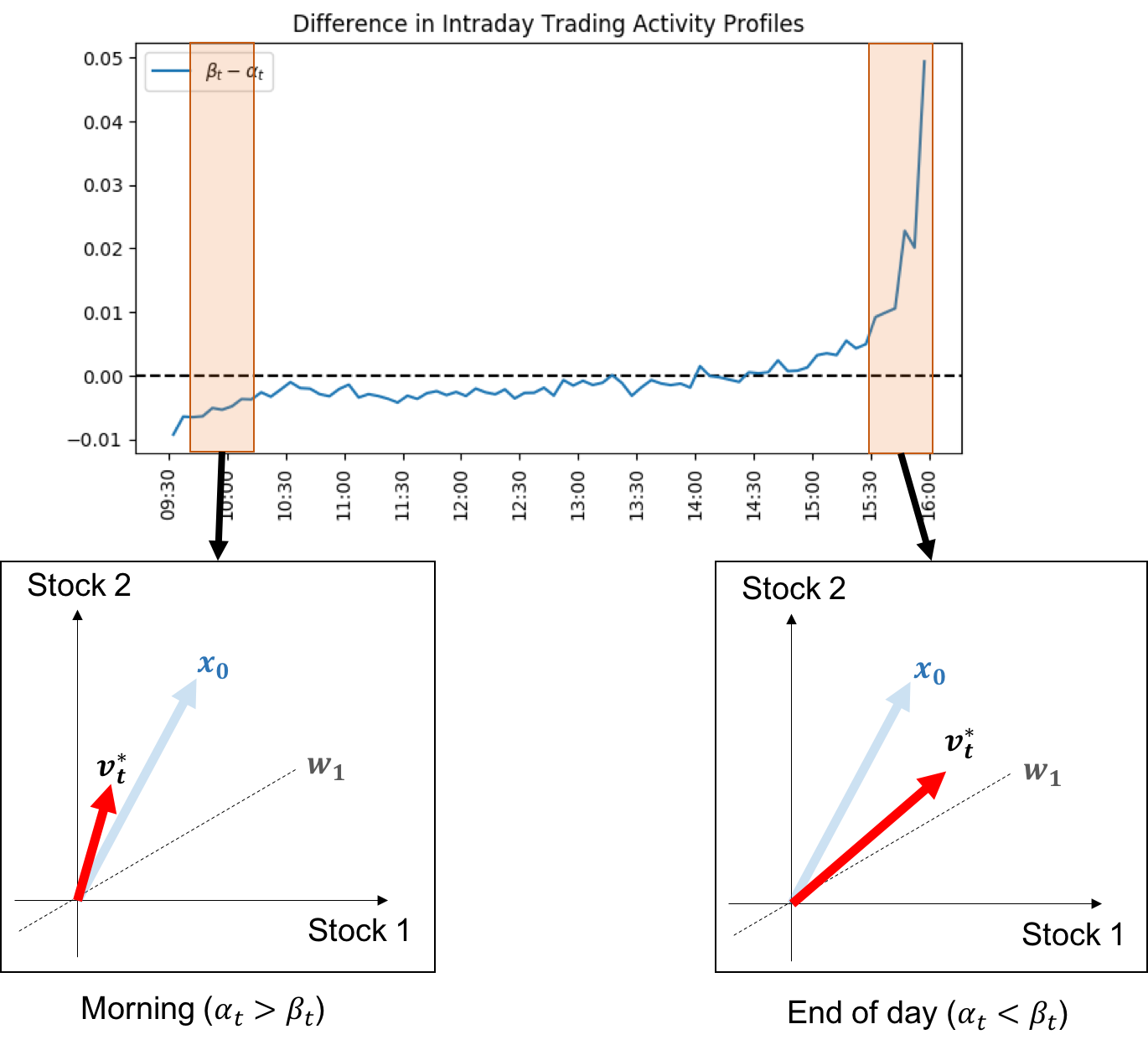} 
	\caption{Illustration of the optimized schedule, which is shown to tilt away or toward the direction of index fund depending on the difference between single stock and index fund liquidity. }
	\label{f-example-opt-exec}
\end{figure}

Finally, Figure \ref{f-example-opt-exec} provides a graphical illustration of the effect of these estimated to the optimal execution schedule in (\ref{e-opt-v}).
The example depicts an investor wants to liquidate a portfolio $\mathbf{x}_0$ with two orders, where the weights of the liquidation portfolio deviate significantly from the weights of the index portfolio $\mathbf{w}_1$ (as captured by the angle between $\mathbf{x}_0$ and $\mathbf{w}_1$).
To exploit the increased end-of-day liquidity in the direction of the index portfolio, $\mathbf{w}_1$, the optimal schedule trades more aggressively stock 2 in the morning session, as shown by $\mathbf{v}_t^*$, thus tilting away from a separable VWAP-like execution that would be aligned with $\mathbf{x}_0$.
As a consequence, the residual portfolio executed towards the end of the day is better aligned with the index portfolio (in the afternoon, $\mathbf{v}_t^*$ is closer to the index portfolio $\mathbf{w}_1$).
%\footnote{For example, one may want to liquidate a portfolio of healthcare stocks, where the sizes of the respective orders differ significantly from the market weights of the healthcare sector. It is plausible to imagine that it is cheaper to source liquidity on a healthcare portfolio that is aligned with the corresponding sector portfolio, as opposed to a portfolio that would be, for example, more concentrated in a few orders and have more idiosyncratic risk.The optimized schedule would tilt away from a separable (VWAP-like) execution to make use of the increased liquidity along the sector portfolio weights that is available towards the end of the day.}

\iffalse
We've made stylized assumptions to parameterize the intraday variation of liquidity and trading volume with profiles $\alpha_t$ and $\beta_t$.
The main purpose of this model is to derive a parsimonious expression for intraday trading volume variation that is quantifiable from the simple statistics as above, and also tractable enough for performance analysis provided in the next section.
By relaxing some of conditions, one may obtain a more realistic model but more complicated, which may require additional statistics to quantify the underlying parameters or may lose tractability.
For the practical purpose, we rather suggest an estimation procedure discussed in \S \ref{ss-estimation}.
\fi

\subsection{Implementation Shortfall Comparison: Optimal vs. Separable Execution Schedules}
\label{ss-perf-bounds}

From \eqref{e-sep-v} and \eqref{e-theta1} we get that the separable schedule
$\mathbf{v}_t^\text{sep}$ is given by:
\begin{equation} \label{e-sep-simple}
\mathbf{v}_t^\text{sep} = \left( \alpha_t \cdot (1-\theta) + \beta_t \cdot \theta \right) \cdot \mathbf{x}_0,
\end{equation}
and for $\mathbf{v}_t^*$ and $\mathbf{v}_t^\text{sep}$, the expected implementation shortfall can be written as follows:
\begin{eqnarray}
\bar{\cal C}(\mathbf{v}_t^*)
&=& \frac{1}{2} \mathbf{x}_0^\top \left( \Pid + \W \Pf \W^\top \right)^{-1} \mathbf{x}_0 , \\
\bar{\cal C}(\mathbf{v}_t^\text{sep})
&=& \frac{1}{2} \sum_{t=1}^T \left( \alpha_t \cdot (1-\theta) + \beta_t \cdot \theta \right)^2 \cdot \mathbf{x}_0^\top  \left( \alpha_t \Pid + \beta_t \W \Pf \W^\top \right)^{-1} \mathbf{x}_0.
\end{eqnarray}
Note that under the assumption that there is only one index fund, $\mathbf{w}_1$, we could simplify the above expressions and reduce $\W \Pf \W^\top$ to
$\mathbf{w}_1 \bar{\psi}_{\fund,1} \mathbf{w}_1^\top$.
The expression for $\bar{\cal C}(\mathbf{v}_t^\text{sep})$ is obtained from 
substituting \eqref{e-sep-simple} into \eqref{e-opt-prob}.
We define as a relative performance measure the ratio between the expected transaction costs incurred by the two execution schedules.
\begin{equation}
\Upsilon(\mathbf{x}_0) \triangleq \frac{ \bar{\cal C}(\mathbf{v}_t^\text{sep}) }{ \bar{\cal C}(\mathbf{v}_t^*)  };
\end{equation}
this ratio is clearly greater or equal to $1$, and captures the additional cost incurred by the separable `VWAP-like' schedule over the optimized, coupled, execution schedule.

\begin{proposition}[Exact cost ratio] For any $\mathbf{x}_0 \in \mathbb{R}^N$,
\label{p-cost-ratio}
	\begin{equation}
	\label{e-cost-ratio}
		\Upsilon(\mathbf{x}_0)
			= 1 + \theta^2 \cdot \left( \sum_{t=1}^T \frac{\beta_t^2}{\alpha_t}-1 \right)
				+  \Delta \cdot \left( \frac{ \mathbf{x}_0^\top \Pid^{-1} \mathbf{x}_0 }{ \left( \mathbf{w}_1^\top \Pid^{-1} \mathbf{x}_0 \right)^2 } \cdot \frac{ 1+\eta_1 }{ \bar{\psi}_{\fund,1} } - 1 \right)^{-1},
	\end{equation}
	where 
	\begin{equation}
	 	\gamma_t \triangleq \frac{\beta_t}{\alpha_t}
		, \quad
		\eta_1 \triangleq \bar{\psi}_{\fund,1} \mathbf{w}_1^\top \Pid^{-1} \mathbf{w}_1
		, \quad  \text{and} \quad
		\Delta \triangleq  \sum_{t=1}^T \frac{ \alpha_t \cdot \left(1 - \theta \cdot (1-\gamma_t) \right)^2 (1-\gamma_t) }{ 1+ \eta_1 \cdot \gamma_t }.
	\end{equation}
\end{proposition}

(The proof is given in Appendix \ref{prf-cost-ratio}.)
The parameter $\eta_1$ is the ratio of index fund liquidity ($1/\bar{\psi}_{\fund,1} $) over the liquidity provided by single stock investors along the index fund weights ($\mathbf{w}_1^\top \Pid^{-1} \mathbf{w}_1$).
Equivalently, it is the ratio of the price change of trading along the index fund direction $\mathbf{w}_1$ against only the portion of single stock investors in the market\footnote{
If we trade $\mathbf{w}_1$ against single stock investors we will cause a change in prices given by $\Delta \mathbf{p} = \Pid^{-1} \mathbf{w}_1$, which would imply a change in the price of the market portfolio equal to $\mathbf{w}_1^\top \Pid^{-1} \mathbf{w}_1$.},
versus the price change of trading along $\mathbf{w}_1$ along only the index fund investors, which is $1/\bar{\psi}_{\fund,1} $.\footnote{
To gain some intuition as to the magnitude of that parameter, imagine wanting to buy a \$100 million slice of the S\&P500, where in one case this is acquired from distinct liquidity providers, each trading only one of constituent orders, while in the other case it is acquired from the same (portfolio) liquidity provider.
The mere difference in the aggregate volatility held by the distinct liquidity providers in the first scenario versus the unique liquidity provider trading the market portfolio in the latter, would suggest a potentially significant difference in trading costs, and therefore a high $( \gg 1)$ value for $\eta_1$. \label{foot-principle}}

The last expression in the performance metric is a product two terms: the first is associated with the intraday variation of liquidity and trading volume ($\Delta$), and the second is associated with the degree of alignment between the execution portfolio $\mathbf{x}_0$ and the index fund weights $\mathbf{w}_1$.
%Roughly speaking, when there is a large discrepancy between single-stock investor's liquidity profile and index-fund investor's liquidity profile (i.e., $\alpha_t$ and $\beta_t$ are following very different trajectories), the former term becomes large, resulting in a large cost reduction.
%In addition, when $\mathbf{x}_0$ is aligned with the index fund $\mathbf{w}_1$, the later term becomes large, so the ratio does.
%We further characterize the cost ratio in the following remarks.

\noindent
\textbf{Worst case liquidation portfolios.}
First, we explore the structure of the portfolios that would exhibit the largest optimality gap under a separable execution.

\begin{remark}[Maximum/minimum cost ratio]
\label{r-minmax}
Let $\Upsilon_\textup{market}$ and $\Upsilon_\textup{orth}$ be the cost ratio when, respectively, $\mathbf{x}_0 = \mathbf{w}_1$, and when $\mathbf{x}_0 = \mathbf{w}_1^\perp$ where $\mathbf{w}_1^\perp$ is an arbitrary portfolio such that $\mathbf{w}_1^\top \Pid^{-1} \mathbf{w}_1^\perp = 0$ with $\mathbf{w}_1^\top \ne \mathbf{0}$.
\begin{eqnarray}
\Upsilon_\textup{market}
&\triangleq& \Upsilon(\mathbf{x}_0 = \mathbf{w}_1)
= 1 + \theta^2 \cdot \left( \sum_{t=1}^T \frac{\beta_t^2}{\alpha_t} - 1 \right) + \eta_1 \cdot \Delta ,
\\
\Upsilon_\textup{orth}
&\triangleq& \Upsilon(\mathbf{x}_0 = \mathbf{w}_1^\perp)
= 1 + \theta^2 \cdot \left( \sum_{t=1}^T \frac{\beta_t^2}{\alpha_t} - 1 \right) .
\end{eqnarray}
Then, largest and smallest cost ratio are obtained at either $\mathbf{x}_0 = \mathbf{w}_1$ or $\mathbf{x}_0 = \mathbf{w}_1^\perp$ depending on the sign of $\Delta$.
\begin{eqnarray}
\max_{\mathbf{x}_0 \in \mathbb{R}^N} \left\{ \Upsilon(\mathbf{x}_0) \right\}
&=& \left\{ \begin{array}{ll} 
	\Upsilon_\textup{market} & \text{if } \Delta \geq 0 \\
	\Upsilon_\textup{orth} & \text{if } \Delta \leq 0 
	\end{array} \right. ,
\\
\min_{\mathbf{x}_0 \in \mathbb{R}^N} \left\{ \Upsilon(\mathbf{x}_0) \right\}
&=& \left\{ \begin{array}{ll} 
	\Upsilon_\textup{orth} & \text{if } \Delta \geq 0 \\
	\Upsilon_\textup{market} & \text{if } \Delta \leq 0 
	\end{array} \right. .
\end{eqnarray}
In particular, for fixed $(\eta_1, \alpha_1, \cdots, \alpha_T, \beta_1, \cdots, \beta_T)$, there exists $\theta^* \in [0,1]$ such that
\begin{equation}
\label{e-sign-delta-theta}
	\Delta \geq 0 \quad \text{if } \theta \leq \theta^*
	\quad \text{and} \quad
	\Delta \leq 0 \quad \text{if } \theta \geq \theta^* .
\end{equation}
%\todo{I will also be able to show that $\Delta$ is a non-increasing function of $\theta$. Include it?}
Similarly, for fixed $(\theta, \alpha_1, \cdots, \alpha_T, \beta_1, \cdots, \beta_T)$, there exists $\eta_1^* \in [\frac{\theta}{1-\theta},\infty]$ such that
\begin{equation}
\label{e-sign-delta-eta}
	\Delta \leq 0 \quad \text{if } \eta_1 \leq \eta_1^*
	\quad \text{and} \quad
	\Delta \geq 0 \quad \text{if } \eta_1 \geq \eta_1^* .
\end{equation}
\end{remark}

This remark identifies which portfolios give rise to the largest and smallest cost ratios, respectively.
It is straightforward that the cost ratio has extreme values at $\mathbf{x}_0 = \mathbf{w}_1$ and $\mathbf{x}_0 = \mathbf{w}_1^\perp$, i.e., when $\mathbf{x}_0$ is most and least aligned with the market portfolio $\mathbf{w}_1$.
%To understand suboptimality of the separable execution in each of these two extreme cases $\mathbf{x}_0 = \mathbf{w}_1$ and $\mathbf{x}_0 = \mathbf{w}_1^\perp$, we need to compare the separable execution and the optimal execution for these two cases:  
From  \eqref{e-sep-simple} and \eqref{e-tilting} we get that in these two extreme cases the separable and optimal schedules are given by:
\[
\mathbf{v}_t^\text{sep} = \left( \alpha_t \cdot (1-\theta) + \beta_t \cdot \theta \right) \cdot \mathbf{x}_0, \quad \text{and} \quad
\mathbf{v}_t^* = \left\{
	\begin{array}{ll}
		\left( \alpha_t \cdot \left( 1 - \frac{\eta_1}{1+\eta_1} \right) + \beta_t \cdot \frac{\eta_1}{1+\eta_1} \right) \cdot \mathbf{x}_0  & \text{if } \mathbf{x}_0 = \mathbf{w}_1, \\
		\alpha_t \cdot \mathbf{x}_0 & \text{if } \mathbf{x}_0 = \mathbf{w}_1^\perp.
	\end{array} \right.
\]
When $\mathbf{x}_0 = \mathbf{w}_1$, the sensitivity of the optimized schedule to the intensity of index-fund liquidity provision, $\beta_t$, is $\frac{\eta_1}{1+\eta_1}$, whereas that of the separable execution is $\theta$.
If $\theta > \frac{\eta_1}{1+\eta_1}$, the separable execution will trade more than is optimal to do in the morning, and trade less than optimal towards the end of the day; 
the opposite happens if $\theta <  \frac{\eta_1}{1+\eta_1}$.
We can expect that the suboptimality of separable execution roughly scales with $\left( \theta - \frac{\eta_1}{1+\eta_1} \right)^2$.
A similar argument suggests that the suboptimality gap when $\mathbf{x}_0 = \mathbf{w}_1^\perp$ roughly scales like $\left( \theta - 0 \right)^2$.
Comparing $\left( \theta - \frac{\eta_1}{1+\eta_1} \right)^2$ and $\theta^2$ as proxies for  $\Upsilon_\text{market}$ and $\Upsilon_\text{orth}$, respectively, leads to the findings of Remark \ref{r-minmax}.

\noindent
\textbf{Performance implications when trading the market portfolio.}
Next we characterize $\Upsilon_\textup{market}$ as a function of the parameter $\eta_1$.

\begin{remark}[Characterization of $\Upsilon_\textup{market}$]
\label{r-upsilon-market}
For fixed $( \theta, \alpha_1, \cdots, \alpha_T, \beta_1, \cdots, \beta_T)$, as a function of $\eta_1$,
\begin{equation}
	\Upsilon_\textup{market}(\eta_1) \text{ decreases if } \eta_1 \leq \frac{\theta}{1-\theta}
	, \quad \text{and} \quad
	\Upsilon_\textup{market}(\eta_1) \text{ increases if } \eta_1 \geq \frac{\theta}{1-\theta} .
\end{equation}
For particular values of $\eta_1$,
\begin{eqnarray}
	\Upsilon_\textup{market}(\eta_1=0) &=& 1 + \theta^2 \cdot \left( \sum_{t=1}^T \frac{\beta_t^2}{\alpha_t} - 1 \right) ,
	\\
	\Upsilon_\textup{market}\left(\eta_1=\frac{\theta}{1-\theta}\right) &=& 1 ,
	\\
\label{e-upper-bound}
	\lim_{\eta_1 \rightarrow \infty} \Upsilon_\textup{market}(\eta_1) &=& 1 + (1-\theta)^2 \cdot \left( \sum_{t=1}^T \frac{\alpha_t^2}{\beta_t} - 1 \right) .
\end{eqnarray}
\end{remark}

It predicts that $\Upsilon_\text{market}$ decreases first and then increases as $\eta_1$ varies.
This can be similarly understood as in Remark \ref{r-minmax}: separable execution correctly reacts to the liquidity provided by index-fund investors only when $\eta_1 = \frac{\theta}{1-\theta}$, and overreacts or underreacts when $\eta_1$ deviates from $\frac{\theta}{1-\theta}$.

\begin{figure}[!h]
	\centering
	\includegraphics[width=0.5\linewidth]{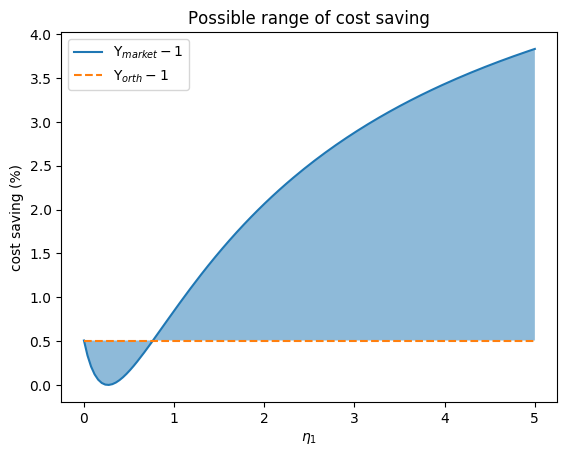}
	\caption{Possible range of cost ratio $\Upsilon$ with respect to $\eta_1$ given the values of $\theta, \alpha_1,\cdots,\alpha_T, \beta_1, \cdots,\beta_T$ obtained in \S \ref{ss-volume-profile}. The coupled execution could save up to 6.2 \% when trading the market portfolio.}
\label{f-cost-ratio}
\end{figure}

Our estimate for the fraction of index fund liquidity, $\theta = .21$, which suggests a threshold value for $\theta/(1-\theta) \approx .27$.
Even though the value of $\eta_1$ is unidentifiable in our context, one would expect the value of $\eta_1$ to be moderately large (cf. Footnote 4), and that the realized benefits from optimizing the coupled execution of the portfolio over that of a separable execution to approach the upper bound in \eqref{e-upper-bound}.
That upper bound is equal to 6.2\% for the parameters $\theta, \alpha_1,\cdots,\alpha_T, \beta_1, \cdots,\beta_T$ estimated in \S \ref{ss-volume-profile}, and Figure \ref{f-cost-ratio} graphs $\Upsilon_\text{market}$ and $\Upsilon_\text{orth}$ as functions of $\eta_1$.% given the values of $\theta, \alpha_1,\cdots,\alpha_T, \beta_1, \cdots,\beta_T$ obtained in \S \ref{ss-volume-profile}.
That is, under the assumptions of our stylized generative model of order flow one could reduce execution costs over the separable `VWAP-like' execution by as much as 6.2\% by optimally coupling the execution schedules of the various orders that are being liquidated so as to exploit the benefits due to portfolio liquidity provision and incorporating cross-impact phenomena.

\noindent
\textbf{Liquidating single orders.}
Finally, we specialize our results to the case where the target portfolio to be liquidated is an order on a single security.

\begin{remark}[Individual orders]
\label{r-individual}
When trading a single stock,
	\begin{equation}
		\Upsilon(\mathbf{x}_0=\mathbf{e}_i) 
			= 1 + \theta^2 \cdot \left( \sum_{t=1}^T \frac{\beta_t^2}{\alpha_t} - 1 \right)
			+  \frac{ \eta_{1,i} }{ 1 + \eta_1 - \eta_{1,i} } \cdot \Delta ,
	\end{equation}
	where $\eta_{1,i} \triangleq \frac{ w_{1i}^2 \cdot \bar{\psi}_{\fund,1} }{ \bar{\psi}_{\idio,i} }$. 
We can further identify the stock that incurs the largest cost ratio:
	\begin{equation} \label{e-single-w}
		\argmax_{i=1,\cdots,N} \left\{ \Upsilon(\mathbf{x}_0=\mathbf{e}_i) \right\}
		= \left\{ \begin{array}{cc}
			\argmax_{i=1,\cdots,N} \left\{ \frac{ w_{1i}^2 }{ \bar{\psi}_{\idio,i} } \right\}
			& \text{if } \Delta \geq 0
		\\
			\argmin_{i=1,\cdots,N} \left\{ \frac{ w_{1i}^2 }{ \bar{\psi}_{\idio,i} } \right\}
			& \text{if } \Delta \leq 0
		\end{array} \right. .
	\end{equation}
\end{remark}

Here we are comparing the performance implications of liquidating a single order via a `VWAP-like' execution versus liquidating it via an optimized schedule that may add positions early in the day, so as to unwind the residual portfolio later in the day in a way that benefits from the liquidity provided by index fund investors.
The fraction $w_{1i}^2 / \bar{\psi}_{\idio,i}$ determines which security is most costly to trade, and depends both on the market weight of the security in the index fund portfolio, and the liquidity provided by its own single-stock investors.
Assuming that for our estimated value for $\theta$, $\eta_1$ is sufficiently large ($\Delta \geq 0$), (\ref{e-single-w}) suggests that the optimized execution schedule may be most beneficial when trading in securities with large market weights. 

\iffalse
Throughout these results, an auxiliary variable $\eta_1$ plays an important role.
Adopting the implications of Proposition \ref{tr_cost}, it can be interpreted as a ratio between the liquidity provided by index-fund investors and the liquidity provided by single-stock investors, particularly along the direction of index fund $\mathbf{w}_1$. 
	\begin{eqnarray}
		\eta_1 & \triangleq & \frac{ \mathbf{w}_1^\top \Pid^{-1} \mathbf{w}_1 }{ \bar{\psi}_{\fund,1}^{-1} }
			\\ &=& \frac{ \text{portfolio price change when executing $\mathbf{w}_1$ against single-stock investors only} }{ \text{portfolio price change when executing $\mathbf{w}_1$ against index-fund investors only} }
			\\ &=& \frac{ \text{``index-fund liquidity along $\mathbf{w}_1$''} }{ \text{``single-stock liquidity along $\mathbf{w}_1$''} }
	\end{eqnarray}
Despite that the value of $\eta_1$ is hard to identify, we guess it could be very large if some stocks are mostly traded by index-fund investors.
\fi

\section{Concluding Remarks} \label{s-practice}

In closing, we briefly review a tractable estimation procedure for the cross-security market impact model, and discuss some practical considerations of portfolio execution algorithms.

%So far we have assumed that the liquidator does not impose any additional trading constraints; this led to a more tractable analysis that served to highlight our key findings in their simplest form -- namely the effect of intraday cross-impact variations in the cost minimizing trade schedule, and the magnitude of these effects on the optimal execution costs.

\subsection{Estimation of Cross-asset Market Impact}
\label{ss-estimation}

Estimating a cross-security impact model that explicitly measures the impact coefficient among any pair of securities $i, j$ is hard due to the high dimensionality of the unknown coefficient matrix ($N \times N$ in this case), and because the underlying data tends to be very noisy.
In this subsection we sketch a procedure that would take as input a large set of proprietary portfolio transactions, and exploit the structure of the cost model postulated in the previous two sections to efficiently estimate an impact cost model with $K+1$ free parameters, where $K$ is the number of index funds along which portfolio investors supply liquidity.

\emph{The data set:}
We assume that the given data contains a set of tuples $(\mathbf{\tilde{v}}_{dt}, \mathbf{\bar{r}}_{dt}, \Wt_d)$,
where $\mathbf{\tilde{v}}_{dt} \in \mathbb{R}^N$ is the portfolio vector executed during time interval $t$ on day $d$, expressed in notional \$ amounts;
$\mathbf{\bar{r}}_{dt} \in \mathbb{R}^N$ is the notional-weighted implementation shortfall incurred in the execution of portfolio $\mathbf{\tilde{v}}_{dt} $ relative to the arrival price vector at the beginning of time interval $t$ on day $d$;
and $\Wt_d \in \mathbb{R}^{N \times K}$ are the dollar-weighted vectors of the $K$ index funds on day $d$. 
In summary, one has access to realized portfolio executions, their realized shortfalls, and reference information about the prevailing weight vectors of popular index funds (such as the market and sector portfolios).

The derivation in \S \ref{s-model} predicts the following relationship between the executed quantity 
$\mathbf{\tilde{v}}_{dt}$ and the realized shortfall $\mathbf{\bar{r}}_{dt}$:\footnote{
In this section, we are using the realized shortfall (return) $\mathbf{\bar{r}}_{dt}$ instead of the absolute price change $\Delta \mathbf{p}_{dt}$, and notional traded vectors $\mathbf{\tilde{v}}_{dt}$ instead of number of shares $\mathbf{v}_{dt}$.
Similarly, we use dollar-weighted vectors $\mathbf{\tilde{w}}_k$ instead of shares-weighted vectors $\mathbf{w}_k$. With the rescaled liquidity parameters $\Pidt$ and $\Pft$, the structure of the price impact model remains the same. 
See Appendix \ref{s-change-of-units}. }
\begin{equation} \label{e-est-general}
\mathbf{\bar{r}}_{dt} = \frac{1}{2} \left( \bm{\tilde{\Psi}}_{\idio,dt} + \Wt_d \bm{\tilde{\Psi}}_{\fund,dt} \Wt_d^\top \right)^{-1} \mathbf{\tilde{v}}_{dt} + \bm{\epsilon}_{dt},
\end{equation}
where the rescaled liquidity matrices are given by
$\bm{\tilde{\Psi}}_{\idio,dt} = \diag (\tilde{\psi}_{\idio,1dt}, \ldots, \tilde{\psi}_{\idio,Ndt})$
and 
$\bm{\tilde{\Psi}}_{\fund,dt} = \diag (\tilde{\psi}_{\fund,1dt} \ldots, \tilde{\psi}_{\fund,Kdt})$. 
Finally, the intraday covariance matrix of the noise term is 
$\widehat{\bm{\Sigma}}_{dt} \triangleq \Cov\left[ \bm{\epsilon}_{dt} \right]$.

The liquidity variable -- or supply function-- $\tilde{\psi}_{\idio,idt}$ (or $\tilde{\psi}_{\fund,kdt}$) represents the notional amount of stock $i$ (or index fund $k$) that will be supplied by single-stock investors (or index fund investors) in response to a movement in the price of the stock or index.  
The liquidity variable $\tilde{\psi}$ captures (i) the number of investors, or participation intensity, present in each period that we are executing, and (ii) their sensitivity to price movements. 
%how much each of them will trade in response to a certain amount of price change. 
The first factor roughly scales in proportion to trading volume, while the second factor varies in a way that depends to the volatility of the underlying security or index, and, specifically, it is plausible to imagine that it scales inversely proportional to the volatility itself.
We will approximate $\tilde{\psi}_{\idio,idt}$'s and $\tilde{\psi}_{\fund,kdt}$'s with the following reduced form parameterizations
\begin{equation}
\tilde{\psi}_{\idio,idt} = \gamma_\idio \cdot \frac{ \widehat{\text{DVol}}_{idt} }{ \widehat{\sigma}_{idt} }, \quad
\tilde{\psi}_{\fund,kdt} = \gamma_{\fund,k} \cdot \frac{ \widehat{\text{DVol}}_{\fund,kdt} }{ \widehat{\sigma}_{\fund,kdt} },
\end{equation}
where $\widehat{\text{DVol}}_{idt}$ and $\widehat{\sigma}_{idt}$ are the forecasted trading volume and the forecasted volatility of stock $i$ on day $d$ at time $t$, and $\gamma_\idio, \gamma_{\fund,1}, \cdots, \gamma_{\fund,K}$ are unknown parameters to be estimated in the sequel.\footnote{We have chosen to use forecasted values for volume and volatility; another option is to use realized values during the execution intervals.
In using a market impact model to make cost predictions of future executions, one has to use forecasted quantities;
e.g., the average trading volume or the realized volatility over the past 30 trading days can be used, possibly after some treatment of outliers.} 
We have selected a simple parametrization where all single stock terms $\tilde{\psi}_{\idio,idt}$ share the same coefficient $\gamma_\idio$ that is believed to reflect some invariant characteristic of all single-stock investors.\footnote{
Such a parameterization of $\tilde{\psi}_{\idio,idt}$ is consistent with most of the literature in estimating market impact models; e.g., assuming that there are only single stock natural liquidity providers, one would recover a commonly encountered cost model of the form $(1/\gamma_\idio)\widehat{\sigma}_{idt} (\texttt{(executed quantity)}/\widehat{\text{DVol}}_{idt})$; cf., \cite{almgren2005direct}).}
The parameterization of $\tilde{\psi}_{\fund,kdt}$ is analogous to that of $\tilde{\psi}_{\idio,idt}$. 
%It contains the (expected) trading volume $\widehat{\text{DVol}}_{\fund,kdt}$  and its (expected) volatility $\widehat{\sigma}_{\fund,kdt}$. 
As a proxy for $\widehat{\text{DVol}}_{\fund,kdt}$ one may use the interval trading volume of a related ETF, a weighted sum of the underlying volume of the constituent stocks, or a projection of intraday market volume along the direction of the index weight vector. 
We allow for distinct coefficients $\gamma_{\fund,k}$ for each index funds, $k$; a more parsimonious parametrization would consider a common coefficient $\gamma_\fund$ for all $k=1, \ldots, K$.

Under this parameterization, \eqref{e-est-general} reduces to
\begin{equation} \label{e-est-spec}
\mathbf{\bar{r}}_{dt} = \frac{1}{2} \mathbf{\tilde{G}}_{dt}\left( \gamma_\idio, \gamma_{\fund,1:K} \right)^{-1} \mathbf{\tilde{v}}_{dt} + \bm{\epsilon}_{dt},
\end{equation}
where the coefficient matrix $\mathbf{\tilde{G}}_{dt}$ takes the form
\begin{equation} \label{e-est-g}
\mathbf{\tilde{G}}_{dt}\left( \gamma_\idio, \gamma_{\fund,1:K} \right)
\triangleq \gamma_\idio \cdot \mathbf{D}_{\idio, dt} + \Wt_d \cdot \mathbf{D}_{\fund, dt} \cdot \Wt_d^\top,
\end{equation}
and
\[
\mathbf{D}_{\idio, dt} = \text{diag}_{i=1}^N\left( \frac{ \widehat{\text{DVol}}_{idt} }{ \widehat{\sigma}_{idt} } \right)
\quad \mbox{and} \quad
\mathbf{D}_{\fund, dt} = \text{diag}_{k=1}^K\left( \gamma_{\fund,k} \cdot \frac{ \widehat{\text{DVol}}_{\fund,kdt} }{ \widehat{\sigma}_{\fund,kdt} } \right).
\]

Given \eqref{e-est-spec}-\eqref{e-est-g} and the observations $(\mathbf{\tilde{v}}_{dt}, \mathbf{\bar{r}}_{dt}, \Wt_d)$ one can pick $\gamma_\idio, \gamma_{\fund,1}, \cdots, \gamma_{\fund,K}$ to best explain the realized execution costs. 
Assuming that the error terms $\bm{\epsilon}_{dt}$ are drawn from multivariate normal distribution with known covariance matrix $\widehat{\bm{\Sigma}}_{dt}$, the log-likelihood is
\[
{\cal L}\left( \gamma_\idio, \gamma_{\fund,1:K} \right) 
= - \sum_{d=1}^D \sum_{t=1}^T \left( \mathbf{\bar{r}}_{dt} -  \frac{1}{2} \mathbf{\tilde{G}}_{dt}\left( \gamma_\idio, \gamma_{\fund,1:K} \right)^{-1} \mathbf{\tilde{v}}_{dt}  \right)^\top \widehat{\bm{\Sigma}}_{dt}^{-1} 
\left( \mathbf{\bar{r}}_{dt} -  \frac{1}{2} \mathbf{\tilde{G}}_{dt}\left( \gamma_\idio, \gamma_{\fund,1:K} \right)^{-1} \mathbf{\tilde{v}}_{dt}  \right).
\]
The parameters $\gamma_\idio, \gamma_{\fund,1}, \cdots, \gamma_{\fund,K}$ can be estimated so as to maximize the log-likelihood ${\cal L}(\cdot)$ (MLE).\footnote{
An alternative approach would use a non-linear regression procedure with an appropriate heteroskedasticity correction.}
The MLE is a non-linear, but low-dimensional, optimization problem; in its implementation, it is useful to again exploit the Woodbury matrix identity in calculating $\mathbf{\tilde{G}}_{dt}^{-1}$, since this can be done by inverting $K \times K$ matrix instead of the larger $N \times N$ matrix.

Given the estimated values of $\gamma_\idio, \gamma_{\fund,1}, \cdots, \gamma_{\fund,K}$, 
one can use the corresponding reduced form impact cost model in the risk-neutral cost minimization problem \eqref{e-opt-prob} to compute the optimal trade schedule, or equivalently plug into the solution in \eqref{e-opt-sol}.

\subsection{Portfolio Execution Algorithms}

\textbf{Trading constraints.}
Trade execution algorithms used to liquidate portfolios may impose additional constraints,
starting with side constraints that force the liquidation schedule to only execute on the securities that are included on the target liquidation portfolio, and to only trade in the direction of the parent orders themselves -- i.e., only sell stock in securities that were submitted as `sell' orders, and vice versa for `buy' orders. 
For example, such side constraints are enforced when investors delegate the execution of a portfolio to a broker-dealer on a so-called `agency' basis, where the broker executes the orders on behalf of its client.
They may also be enforced in asset management firms where portfolio construction and trade execution are treated as separate functions, and execution traders do not have discretion to deviate from the target liquidation portfolio. 

\S \ref{s-execution} - \S \ref{s-illustration} did not impose these side trading constraints, and the derived optimal schedules may violate this restrictions, e.g., choosing to trade in securities that are outside those in the target liquidation portfolio, $\mathbf{x}_0$, so that the residual liquidation portfolio towards the end of the day could take advantage of (cheaper) portfolio natural liquidity.\footnote{In a market where all liquidity is provided from single stock investors, the optimal schedule would never choose to trade outside the universe of securities that were in the liquidation portfolio or to trade against the direction of the respective parent orders.}
Similarly, the optimal schedule may choose to increase the size of an existing order (as opposed to start liquidating it) early in the day, if that would be beneficial when liquidating the residual portfolio towards the end of the day. 

The constrained portfolio liquidation problem is similar in nature to the one studied in the previous section, and (the numerically) optimized schedule will continue to incorporate and exploit the effect of cross-impact and natural portfolio liquidity provision.
One exception is when liquidating a single parent order, in which case these cross-impact and portfolio liquidity factors are not relevant in such a constrained formulation.\footnote{
In practice, the liquidation problem may impose additional trading constraints, e.g., upper bounds on the speed of execution, (linear) exposure constraints, etc. 
In addition, the liquidator may either incorporate a risk term in her objective function, or add a risk budget constraint.
The resulting problem continues to be a convex quadratic program with similar structural properties.}

\textbf{Functional form of the impact cost model.}
In \S \ref{s-model} we made two assumptions: a) natural liquidity is supplied in quantities that scale linearly as a function of the price dislocation; and b) that the price impact is transient.
These two assumptions allowed us to proceed with an intuitive, closed form analysis, and extract insights on the structure of the optimal trade schedules for a risk neutral liquidator.
The key structural findings of our analysis, however, remain valid even if we consider a model of liquidity provision that leads to sub-linear impact or a model that incorporates impact decay.

\iffalse
\subsection{Optimal execution in practice}

In a consideration of dynamic execution, the execution schedule could be updated adaptively. As execution proceeds, the predicted trajectories of trading volume and volatility would be updated and the schedule can be adjusted accordingly by re-solving the optimization problem for the remaining time horizon.

One can also make a quick modification on the formulation of optimization problem \eqref{e-opt-prob} to obtain more realistic schedule. A constraint on execution schedule $\mathbf{v}_1, \cdots, \mathbf{v}_T$ can be added to exclude the unwanted stocks or to fix the trading direction of individual stocks along the execution horizon. Alternatively, the spreads $\bm{\delta} \in \mathbb{R}_+^N$ can be included in the objective function so that penalizes buying and selling the same stock alternately.
\fi

\bibliography{factor-volume}

\newpage
\appendix
{\small

\section{Change of Units} \label{s-change-of-units}

In \S \ref{s-model} and \S \ref{s-execution}, the price impact was the equilibrium expected price change $\Delta \mathbf{p}$, expressed in dollars, required for the market to clear when executing a vector $\mathbf{v}$, expressed in number of shares for each security in the executed portfolio. 
%There units might be impractical in that its value could be inconsistent before and after an event like stock split. 
%Instead of using the absolute price change in \$ and the executed quantities in shares, 
We can restate the market impact in terms of the return $\mathbf{r} \in \mathbb{R}^N$ as a function of the vector of notional execution quantities $\mathbf{\tilde{v}} \in \mathbb{R}^N$. 
Let $p$ denote the (arrival) equilibrium price vector $\mathbf{p} \in \mathbb{R}^N$, snapped at the beginning of the execution period, and define the diagonal matrix $\mathbf{P} \triangleq \text{diag}(\mathbf{p}) \in \mathbb{R}^{N \times N}$.
Then, 
\begin{equation}
\mathbf{r} \triangleq \mathbf{P}^{-1} \Delta \mathbf{p} \quad \mbox{and} \quad
\mathbf{\tilde{v}} \triangleq \mathbf{P} \mathbf{v}.
\end{equation}
We redefine the liquidity variable $\psi_{\idio,i}$, $\psi_{\fund,k}$ and weight vectors $\mathbf{w}_k$ accordingly:
\begin{equation}
\tilde{\psi}_{\idio,i} \triangleq p_{i}^2 \cdot \psi_{\idio,i}, \quad
\tilde{\psi}_{\fund,k} \triangleq  (\mathbf{w}_k^\top \mathbf{p} )^2 \cdot \psi_{\fund,k} 
\quad \mbox{and} \quad
\mathbf{\tilde{w}}_k \triangleq \frac{ \mathbf{P} \mathbf{w}_k }{ \mathbf{p}^\top \mathbf{w}_k }.
\end{equation}
The redefined liquidity variable $\psi_{\idio,i}$ now has the following interpretation: 
single-stock investors will sell (or buy) $1\% \cdot \tilde{\psi}_{\idio,it}$ dollar amount of stock $i$, when its price rises (or drops) by 1 \%. 
The rescaled weight vector $\mathbf{\tilde{w}}_k$ represents the normalized dollar-weighted portfolio. 
%The cross-sectional market impact can be represented with these new parameters while keeping its structure.
Putting it all together
\begin{eqnarray}
\mathbf{r} 
&=& \mathbf{P}^{-1} \mathbf{G} \mathbf{v}
= \mathbf{P}^{-1} \left( \bm{\Psi}_\idio + \W \bm{\Psi}_\fund \W^\top \right)^{-1} \mathbf{P}^{-1} \cdot \mathbf{P} \mathbf{v} \nonumber \\
&=& \left( \mathbf{P} \bm{\Psi}_\idio \mathbf{P} + \mathbf{P} \W \bm{\Psi}_\fund \W^\top \mathbf{P} \right)^{-1} \mathbf{\tilde{v}} 
= \left( \Pidt + \Wt \Pft \Wt^\top \right)^{-1} \mathbf{\tilde{v}}. \nonumber
\end{eqnarray}
The resulting expected implementation shortfall cost is unchanged:
\[
\bar{ {\cal C} }(\mathbf{v}) \triangleq \frac{1}{2} \mathbf{v}^\top \Delta \mathbf{p} 
= \frac{1}{2} \mathbf{\tilde{v}}^\top \mathbf{r}
= \frac{1}{2} \mathbf{\tilde{v}}^\top \left( \Pidt + \Wt \Pft \Wt^\top \right)^{-1} \mathbf{\tilde{v}}.
\]

\section{Proofs}

\subsection{Proof of Proposition \ref{p-tr-cost}} \label{prf-tr-cost}

\noindent
We first focus on the case where $\mathbf{v} = \mathbf{W} \mathbf{u}$ in \eqref{e-tr-cost-fact}. 
When $\alpha = \beta = 1$, by Woodbury identity we get that
\[
\mathbf{G} 
= \left( \bm{\Psi}_{\idio} + \W \bm{\Psi}_{\fund} \W^\top \right)^{-1}
= \bm{\Psi}_{\idio}^{-1} - \bm{\Psi}_{\idio}^{-1} \W \left( \bm{\Psi}_{\fund}^{-1} + \W^\top \bm{\Psi}_{\idio}^{-1} \W \right)^{-1} \W^\top \bm{\Psi}_{\idio}^{-1} .
\]
Consequently,
\begin{eqnarray*}
\W^\top \mathbf{G} \W 
&=& \W^\top \bm{\Psi}_{\idio}^{-1} \W - \W^\top \bm{\Psi}_{\idio}^{-1} \W \left( \bm{\Psi}_{\fund}^{-1} + \W^\top \bm{\Psi}_{\idio}^{-1} \W \right)^{-1} \W^\top \bm{\Psi}_{\idio}^{-1} \W
\\ &=& \left( \left( \W^\top \bm{\Psi}_{\idio}^{-1} \W \right)^{-1} + \bm{\Psi}_{\fund}  \right)^{-1}.
\end{eqnarray*}
Next, we incorporate the effect of $\alpha$ and $\beta$ as follows:
\[
\W^\top \mathbf{G} \W
= \left( \alpha \cdot \left( \W^\top \bm{\Psi}_{\idio}^{-1} \W \right)^{-1} + \beta \cdot \bm{\Psi}_{\fund}  \right)^{-1} 
\longrightarrow \bm{\Psi}_{\fund}^{-1} 
\quad \text{as } \alpha \rightarrow 0 \text{ and } \beta \rightarrow 1.
\]
Therefore, for any $\mathbf{u} \in \mathbb{R}^K$,
\[
\lim_{\alpha \rightarrow 0, \beta \rightarrow 1} \bar{\cal C}\left( \mathbf{v} = \W \mathbf{u} \right) 
= \lim_{\alpha \rightarrow 0, \beta \rightarrow 1} \frac{1}{2} \mathbf{u}^\top \W^\top \mathbf{G} \W \mathbf{u}
= \frac{1}{2} \mathbf{u}^\top \bm{\Psi}_{\fund}^{-1} \mathbf{u}.
\]

Next we consider the case when $\mathbf{v} \notin \text{span}(\mathbf{w}_1,\cdots,\mathbf{w}_K)$. 
Let $\mathbf{v} = \W \mathbf{u} + \mathbf{e}$ for some $\mathbf{e} \in \mathbb{R}^N$ such that $\W^\top \mathbf{e} = \mathbf{0}$ and $\mathbf{e} \ne \mathbf{0}$. 
Using Woodbury matrix identity,
\[
\mathbf{G}
= \alpha^{-1} \cdot \bm{\Psi}_{\idio}^{-1}  - \alpha^{-1} \cdot \bm{\Psi}_{\idio}^{-1} \W \left( \frac{\alpha}{\beta} \bm{\Psi}_{\fund}^{-1} + \W^\top \bm{\Psi}_{\idio}^{-1} \W \right)^{-1} \W^\top \bm{\Psi}_{\idio}^{-1} .
\]
Therefore,
\[
\lim_{\alpha \rightarrow 0, \beta \rightarrow 1} \left\{ \alpha \cdot \mathbf{G} \right\}
= \bm{\Psi}_{\idio}^{-1}  - \bm{\Psi}_{\idio}^{-1} \W \left( \W^\top \bm{\Psi}_{\idio}^{-1} \W \right)^{-1} \W^\top \bm{\Psi}_{\idio}^{-1}.
\]
With $\mathbf{r} \triangleq \bm{\Psi}_{\idio}^{-1/2} \mathbf{e}$ and $\mathbf{A} \triangleq \bm{\Psi}_{\idio}^{-1/2} \W$,
\[
\mathbf{e}^\top \left( \bm{\Psi}_{\idio}^{-1}  - \bm{\Psi}_{\idio}^{-1} \W \left( \W^\top \bm{\Psi}_{\idio}^{-1} \W \right)^{-1} \W^\top \bm{\Psi}_{\idio}^{-1} \right) \mathbf{e}
= \mathbf{r}^\top \mathbf{r} - \mathbf{r}^\top \mathbf{A} \left( \mathbf{A}^\top \mathbf{A} \right)^{-1} \mathbf{A}^\top \mathbf{r}.
\]
Note that $\mathbf{A} \left( \mathbf{A}^\top \mathbf{A} \right)^{-1} \mathbf{A}^\top \mathbf{r}$ is a projection of $\mathbf{r}$ onto the space spanned by $\mathbf{A}$ (denoted by $\text{span}(\mathbf{A})$). 
Therefore,
\[
\lim_{\alpha \rightarrow 0, \beta \rightarrow 1} \left\{ \alpha \cdot \mathbf{e}^\top \mathbf{G} \mathbf{e} \right\} = 0
\quad \text{if and only if} \quad
\mathbf{r} \in \text{span}(\mathbf{A}).
\]
If $\mathbf{r} \in \text{span}(\mathbf{A})$: i.e., $\mathbf{r} = \mathbf{A}\mathbf{s}$ for some $\mathbf{s} \in \mathbb{R}^K$, $\mathbf{e} = \bm{\Psi}_{\idio}^{1/2} \mathbf{r} = \bm{\Psi}_{\idio}^{1/2} \bm{\Psi}_{\idio}^{-1/2} \W \mathbf{s} = \W \mathbf{s}$, and hence $\mathbf{v} \in \text{span}(\W)$. 
Since we are assuming $\mathbf{v} \notin \text{span}(\W)$, we have $\mathbf{r} \notin \text{span}(\mathbf{A})$, and therefore
\[
\lim_{\alpha \rightarrow 0, \beta \rightarrow 1} \left\{ \alpha \cdot \mathbf{e}^\top \mathbf{G} \mathbf{e} \right\} > 0.
\]
Furthermore,
\begin{eqnarray}
\mathbf{G} \W
&=& \alpha^{-1} \cdot \bm{\Psi}_{\idio}^{-1} \W  - \alpha^{-1} \cdot \bm{\Psi}_{\idio}^{-1} \W \left( \frac{\alpha}{\beta} \bm{\Psi}_{\fund}^{-1} + \W^\top \bm{\Psi}_{\idio}^{-1} \W \right)^{-1} \W^\top \bm{\Psi}_{\idio}^{-1} \W
\nonumber \\ 
&=& \alpha^{-1} \cdot \bm{\Psi}_{\idio}^{-1} \W \underbrace{ \left( \mathbf{I}_K - \left[ \frac{\alpha}{\beta} \left(\W^\top \bm{\Psi}_{\idio}^{-1} \W \right)^{-1} \bm{\Psi}_{\fund}^{-1} + \mathbf{I}_K \right]^{-1} \right) }_{\longrightarrow \mathbf{O} \text{ as } \alpha \rightarrow 0} . \nonumber
\end{eqnarray}
Therefore,
\[
\lim_{\alpha \rightarrow 0, \beta \rightarrow 1} \left\{ \alpha \cdot \mathbf{e}^\top \mathbf{G} \W \mathbf{u} \right\} = 0.
\]
To summarize, together with the fact that $\lim_{\alpha \rightarrow 0, \beta \rightarrow 1} \left\{ \mathbf{u} \W^\top \mathbf{G} \W \mathbf{u} \right\} = \mathbf{u}^\top \bm{\Psi}_{\fund}^{-1} \mathbf{u}$,
\begin{eqnarray}
&& \lim_{\alpha \rightarrow 0, \beta \rightarrow 1} \left\{ \alpha \cdot \left( \W \mathbf{u} +  \mathbf{e} \right)^\top \mathbf{G} \left( \W \mathbf{u} +  \mathbf{e} \right) \right\}
\nonumber \\ 
&=& \lim_{\alpha \rightarrow 0, \beta \rightarrow 1} \left\{ \alpha \cdot \mathbf{u} \W^\top \mathbf{G} \W \mathbf{u} \right\} 
+ \lim_{\alpha \rightarrow 0, \beta \rightarrow 1} \left\{ \alpha \cdot 2 \mathbf{e}^\top \mathbf{G} \W \mathbf{u} \right\} 
+ \lim_{\alpha \rightarrow 0, \beta \rightarrow 1} \left\{ \alpha \cdot \mathbf{e}^\top \mathbf{G} \mathbf{e} \right\}
\nonumber \\ 
&=& 0 + 0 + \lim_{\alpha \rightarrow 0, \beta \rightarrow 1} \left\{ \alpha \cdot \mathbf{e}^\top \mathbf{G} \mathbf{e} \right\}  > 0.
\end{eqnarray}
It follows that 
$\lim_{\alpha \rightarrow 0, \beta \rightarrow 1} \left\{ \left( \W \mathbf{u} +  \mathbf{e} \right)^\top \mathbf{G} \left( \W \mathbf{u} +  \mathbf{e} \right) \right\}= \infty$.
\qed
	
\subsection{Proof of Proposition \ref{p-tilting}}  \label{prf-tilting}
	
Note that
\begin{eqnarray}
&& \W \Pf \W^\top \Pid^{-1} - \W \Pf \W^\top \Pid^{-1} \W \left( \Pf^{-1} + \W^\top \Pid^{-1} \W \right)^{-1} \W^\top \Pid^{-1}
\nonumber \\ 
&=& \W \Pf \cdot \left( \Pf^{-1} + \W^\top \Pid^{-1} \W \right) \cdot \left( \Pf^{-1} + \W^\top \Pid^{-1} \W  \right)^{-1} \W^\top \Pid^{-1} 
\nonumber \\ 
&& - \W \Pf \cdot \W^\top \Pid^{-1} \W \cdot \left( \Pf^{-1} + \W^\top \Pid^{-1} \W  \right)^{-1} \W^\top \Pid^{-1}
\nonumber \\ 
&=& \W \Pf \cdot \left( \Pf^{-1} + \W^\top \Pid^{-1} \W - \W^\top \Pid^{-1} \W \right) \cdot \left( \Pf^{-1} + \W^\top \Pid^{-1} \W  \right)^{-1} \W^\top \Pid^{-1}
\nonumber \\ 
&=& \W  \left( \Pf^{-1} + \W^\top \Pid^{-1} \W  \right)^{-1} \W^\top \Pid^{-1}.
\nonumber 
\end{eqnarray}
Again using the Woodbury matrix identity,
\begin{eqnarray}
%&& \mathbf{G}_t^{-1} \left( \sum_{s=1}^T \mathbf{G}_s^{-1} \right)^{-1} \\
\mathbf{G}_t^{-1} \left( \sum_{s=1}^T \mathbf{G}_s^{-1} \right)^{-1} 
&=& \left( \alpha_t \Pid + \beta_t \W \Pf \W^\top \right) \left( \Pid + \W \Pf \W^\top \right)^{-1} 
\nonumber \\ 
&=& \left( \alpha_t \Pid + \beta_t \W \Pf \W^\top \right) \left( \Pid^{-1} - \Pid^{-1} \W \left( \Pf^{-1} + \W^\top \Pid^{-1} \W \right)^{-1} \W^\top \Pid^{-1} \right)
\nonumber \\ 
&=& \alpha_t \mathbf{I}_N 
- \alpha_t \W \left( \Pf^{-1} + \W^\top \Pid^{-1} \W \right)^{-1} \W^\top \Pid^{-1}
\nonumber \\ 
&& + \beta_t \W \Pf \W^\top \Pid^{-1}
- \beta_t \W \Pf \W^\top \Pid^{-1} \W \left( \Pid^{-1} + \W^\top \Pid^{-1} \W \right)^{-1} \W^\top \Pid^{-1}
%\\ &=& \alpha_t \mathbf{I}_N 
%		- \alpha_t \W \left( \Pf^{-1} + \W^\top \Pid^{-1} \W \right)^{-1} \W^\top \Pid^{-1}
%		+ \beta_t \W \left( \Pf^{-1} + \W^\top \Pid^{-1} \W \right)^{-1} \W^\top \Pid^{-1}
\nonumber \\ 
&=& \alpha_t \mathbf{I}_N + (\beta_t - \alpha_t) \W \underbrace{ \left( \Pf^{-1} + \W^\top \Pid^{-1} \W  \right)^{-1} \W^\top \Pid^{-1} }_{\triangleq \mathbf{\widehat{W}}^\top }. \nonumber
\end{eqnarray}
\qed

\subsection{Simple Generative Model of Order Flow used in \S \ref{s-illustration}}
\label{app-generative}

We first establish an explicit relationship between intraday variation of natural liquidity and intraday variation of the resulting traded volume by introducing a stochastic process generative model for trading volume.
The underlying motivation is simple yet intuitive: single-stock and index-fund investors create (stochastic) order flows onto the securities they wish to trade.
The arrival intensity of these order flows per type of investor in each time period is proportional to the corresponding trading activity or liquidity provided by this investor type in this time period.
This is captured by the profiles $\alpha_t$ and $\beta_t$, respectively.
%) commonly governs the intraday variation of trading activity, liquidity and traded volume attributable to single-stock investors (or index-fund investors).

Specifically, we assume that the notional trade volume of stock $i$ on day $d$ in time interval $t$, $\text{DVol}_{idt}$, is composed of order flows made by single-stock investors $Q_{\idio,idt}$ and a $|\tilde{w}_{1i}|$ proportion of order flows made by index-fund investors $Q_{\fund,dt}$.
$|\tilde{w}_{1i}|$ represents dollar-weighted ownership of stock $i$ in the index fund so that trading one dollar amount of index fund accumulates $|\tilde{w}_{1i}|$ dollar amount of notional trade volume onto stock $i$.
Each order flow can naturally be decomposed into small transactions.
\begin{equation}
\text{DVol}_{idt} = Q_{\idio,idt} + |\tilde{w}_{1i}| \cdot Q_{\fund,dt} = \sum_{j=1}^{N_{\idio,idt}} q_{\idio,idt}(j) + |\tilde{w}_{1i}| \cdot \sum_{j=1}^{N_{\fund,dt}} q_{\fund,dt}(j),
\end{equation}
where $N_{\idio,idt}$ and $q_{\idio,idt}(j)$ represent \# of transactions and the absolute size of the $j^{th}$ transaction made by single-stock investors on day $d$ in time interval $t$. $N_{\fund,dt}$ and $q_{\fund,dt}(j)$ are defined analogously.
We treat $N_{\idio,idt}$, $N_{\fund,dt}$, $q_{\idio,idt}(j)$ and $q_{\fund,dt}(j)$ as random variables that follow particular distribution assumptions.

The order arrival processes for the two investor types are assumed to be Poisson with time-varying rates that are proportional to $\alpha_t$ and $\beta_t$. 
%The time-varying rates capture the fact that when the investors participate the market actively, they will provide abundant liquidity and create large order flows accordingly.
\begin{equation}
N_{\idio,idt} \sim \text{Poisson}( \alpha_t \cdot \Lambda )
\quad \text{and} \quad
N_{\fund,dt} \sim \text{Poisson}( \beta_t \cdot \Lambda ).
\end{equation}
We further assume that the individual order quantities $q_{\idio,idt}(j)$'s (and $q_{\fund,dt}(j)$'s) are all independent and identically distributed with following moment conditions.
\begin{equation}
\E\left[ q_{\idio,idt}(j) \right] = \bar{q}_{\idio,i}, \quad \Var\left[ q_{\idio,idt}(j) \right] = c_v^2 \cdot \bar{q}_{\idio,i}^2, \quad
\E\left[ q_{\fund,dt}(j) \right] = \bar{q}_\fund, \quad \Var\left[ q_{\idio,idt}(j) \right] = c_v^2 \cdot \bar{q}_\fund^2,
\end{equation}
where $c_v$ represents a coefficient of variation.

Under the above assumptions, the single-stock investor's order flow $Q_{\idio,idt}$ is a compound Poisson process with following mean and variance:
\begin{eqnarray}
\E\left[ Q_{\idio,idt} \right] &=& \E\left[ N_{\idio,idt} \right] \cdot \E\left[ q_{\idio,idt}(j) \right] = \alpha_t \cdot \Lambda \cdot \bar{q}_{\idio,i}
\\
\Var\left[ Q_{\idio,idt} \right] &=& \E\left[ \Var\left( Q_{\idio,idt} | N_{\idio,idt} \right) \right] + \Var\left[ \E\left( Q_{\idio,idt} | N_{\idio,idt} \right) \right]
\\ &=& \E\left[ N_{\idio,idt} \cdot c_v^2 \cdot \bar{q}_{\idio,i}^2 \right] + \Var\left[ N_{\idio,idt} \cdot \bar{q}_{\idio,i} \right]
\\ &=& \alpha_t \cdot \Lambda \cdot (c_v^2+1) \cdot \bar{q}_{\idio,i}^2.
\end{eqnarray}
The mean and variance of $Q_{\fund,dt}$ can be expressed in a similar manner.
Summing these flows for each security we get that
\begin{eqnarray}
\E\left[ \text{DVol}_{idt} \right] &=& \alpha_t \cdot \Lambda \cdot \bar{q}_{\idio,i} + \beta_t \cdot \Lambda \cdot |\tilde{w}_{1i}| \cdot \bar{q}_\fund
\\
\Var\left[ \text{DVol}_{idt} \right] &=& \alpha_t \cdot \Lambda \cdot (1+c_v^2) \cdot \bar{q}_{\idio,i}^2 + \beta_t \cdot \Lambda \cdot |\tilde{w}_{1i}|^2 \cdot (1+c_v^2) \cdot \bar{q}_\fund^2
\\
\Cov\left[ \text{DVol}_{idt}, \text{DVol}_{jdt} \right] &=&  \beta_t \cdot \Lambda \cdot |\tilde{w}_{1i}| \cdot |\tilde{w}_{1j}| \cdot (1+c_v^2) \cdot \bar{q}_\fund^2.
\end{eqnarray}
The common order flow $Q_{\fund,dt}$ made by index-fund investors, results into a positive correlation between stocks represented in the index.

Define $\theta_i$ to be the proportion of daily traded volume generated by index-fund investors out of the total daily traded volume of stock $i$.
\begin{equation}
\theta_i \triangleq \frac{ \sum_{t=1}^T \E\left[ |\tilde{w}_{1i}| \cdot Q_{\fund,dt} \right] }{ \sum_{t=1}^T \E\left[ \text{DVol}_{idt} \right] }
	= \frac{ |\tilde{w}_{1i}| \cdot \bar{q}_\fund }{ \bar{q}_{\idio,i} + |\tilde{w}_{1i}| \cdot \bar{q}_\fund }.
\end{equation}
The intraday traded volume profile $\text{VolAlloc}_{it}$ and the pairwise correlation $\text{Correl}_{ijt}$, defined in \eqref{e-vol-alloc-def} and \eqref{e-pair-correl-def}, can be simply expressed with $\theta_i$ and $\theta_j$.
\begin{eqnarray}
\label{e-vol-alloc}
\text{VolAlloc}_{it} &\equiv& \frac{ \E\left[ \text{DVol}_{idt} \right] }{ \sum_{s=1}^T \E\left[ \text{DVol}_{ids} \right] }
	= \alpha_t \cdot (1-\theta_i) + \beta_t \cdot \theta_i
\\
\text{Correl}_{ijt} &\equiv& \frac{ \Cov\left[ \text{DVol}_{idt}, \text{DVol}_{jdt} \right] }{ \sqrt{\Var\left[ \text{DVol}_{idt} \right] } \cdot \sqrt{ \Var\left[ \text{DVol}_{jdt} \right] } }
\\ &=& \frac{ \beta_t \cdot \theta_i \cdot \theta_j }{ \sqrt{ \alpha_t \cdot (1-\theta_i)^2 + \beta_t \cdot \theta_i^2 } \cdot \sqrt{ \alpha_t \cdot (1-\theta_j)^2 + \beta_t \cdot \theta_j^2 }  }
\end{eqnarray}
If we further assume that the proportions $\theta_i$ are the same across all securities,
\begin{equation}
\label{e-theta}
	\theta \equiv \theta_1 = \theta_2 = \cdots = \theta_N,
\end{equation}
then $\text{VolAlloc}_{it}$ is the same for all stocks $i$ and $\text{Correl}_{ijt}$ is identical across all pairs of stocks, $i, j$, as given in \eqref{e-theta1}-\eqref{e-theta2}.
%Even though this setting is quite strong and may not be realistic, it allows parsimonious representations of $\text{AvgVolAlloc}_t$ and $\text{AvgCorrel}_t$ as below.
\iffalse
Specifically, 
\begin{eqnarray*}
%\label{e-avg-vol-alloc}
\text{AvgVolAlloc}_t &=& \frac{1}{N} \sum_{i=1}^N \frac{ \E\left[ \text{DVol}_{idt} \right] }{ \sum_{s=1}^T \E\left[ \text{DVol}_{ids} \right] } = \alpha_t \cdot (1-\theta) + \beta_t \cdot \theta
\\
\text{AvgCorrel}_t &=& \frac{1}{N(N-1)} \sum_{i \ne j} \text{Correl}_{ijt} = \frac{\beta_t \cdot \theta^2}{\alpha_t \cdot (1-\theta)^2 + \beta_t \cdot \theta^2}.
\end{eqnarray*}
Given the actual values of $\text{AvgVolAlloc}_t$ and $\text{AvgCorrel}_t$ visualized in figure \ref{f-empirical}, by solving systems of equations, we can exactly identify the values of $\theta$, $\alpha_1, \cdots, \alpha_T$ and $\beta_1, \cdots, \beta_T$.

\fi

\subsection{Proofs for \S \ref{ss-perf-bounds}}

\subsubsection{Proof of Proposition \ref{p-cost-ratio}}
\label{prf-cost-ratio}

Note that
\begin{eqnarray*}
	\Upsilon(\mathbf{x}_0)
		&=& \frac{ \sum_{t=1}^T (\alpha_t \cdot (1-\theta) + \beta_t \cdot \theta)^2 \cdot \mathbf{x}_0^\top \left( \alpha_t \Pid + \beta_t \W \Pf \W^\top \right)^{-1} \mathbf{x}_0 }{ \mathbf{x}_0^\top \left( \Pid + \W \Pf \W^\top \right)^{-1} \mathbf{x}_0 }
		\\&=& \sum_{t=1}^T \alpha_t \cdot \left(1 + \theta \cdot (\gamma_t-1) \right)^2 \cdot  \frac{ \mathbf{x}_0^\top \left( \Pid + \gamma_t \W \Pf \W^\top \right)^{-1} \mathbf{x}_0 }{ \mathbf{x}_0^\top \left( \Pid + \W \Pf \W^\top \right)^{-1} \mathbf{x}_0 }.
\end{eqnarray*}
By Woodbury's matrix identity,
	\begin{eqnarray*}
		&& \frac{ \mathbf{x}_0^\top \left( \Pid + \gamma_t \W \Pf \W^\top \right)^{-1} \mathbf{x}_0 }{ \mathbf{x}_0^\top \left( \Pid + \W \Pf \W^\top \right)^{-1} \mathbf{x}_0 }
			\\ &=& \frac{ \mathbf{x}_0^\top \Pid^{-1} \mathbf{x}_0 - \mathbf{x}_0^\top \Pid^{-1} \W \left( \gamma_t^{-1} \Pf^{-1} + \W^\top \Pid^{-1} \W \right)^{-1} \W^\top \Pid^{-1} \mathbf{x}_0 }{ 
				\mathbf{x}_0^\top \Pid^{-1} \mathbf{x}_0 - \mathbf{x}_0^\top \Pid^{-1} \W \left( \Pf^{-1} + \W^\top \Pid^{-1} \W \right)^{-1} \W^\top \Pid^{-1} \mathbf{x}_0 }
			\\ &=& 1 + \frac{ \left( \mathbf{x}_0^\top \Pid^{-1} \W \left( \Pf^{-1} + \W^\top \Pid^{-1} \W \right)^{-1} \W^\top \Pid^{-1} \mathbf{x}_0 \right) - \left( \mathbf{x}_0^\top \Pid^{-1} \W \left( \gamma_t^{-1} \Pf^{-1} + \W^\top \Pid^{-1} \W \right)^{-1} \W^\top \Pid^{-1} \mathbf{x}_0 \right) }{ 
				\mathbf{x}_0^\top \Pid^{-1} \mathbf{x}_0 - \mathbf{x}_0^\top \Pid^{-1} \W \left( \Pf^{-1} + \W^\top \Pid^{-1} \W \right)^{-1} \W^\top \Pid^{-1} \mathbf{x}_0 }
			\\ &=& 1 + \frac{  \mathbf{x}_0^\top \Pid^{-1} \W \left( \left( \Pf^{-1} + \W^\top \Pid^{-1} \W \right)^{-1} - \left( \gamma_t^{-1} \Pf^{-1} + \W^\top \Pid^{-1} \W \right)^{-1} \right) \W^\top \Pid^{-1} \mathbf{x}_0 }{ 
				\mathbf{x}_0^\top \Pid^{-1} \mathbf{x}_0 - \mathbf{x}_0^\top \Pid^{-1} \W \left( \Pf^{-1} + \W^\top \Pid^{-1} \W \right)^{-1} \W^\top \Pid^{-1} \mathbf{x}_0 }
			\\ &=& 1 + \frac{  \mathbf{x}_0^\top \Pid^{-1} \W \left(\Pf^{-1} + \W^\top \Pid^{-1} \W \right)^{-1} \cdot \left( \gamma_t^{-1} -1 \right) \Pf^{-1} \cdot \left( \gamma_t^{-1} \Pf^{-1} + \W^\top \Pid^{-1} \W \right)^{-1} \W^\top \Pid^{-1} \mathbf{x}_0 }{
				\mathbf{x}_0^\top \left( \Pid + \W \Pf \W^\top \right)^{-1} \mathbf{x}_0 }
			\\ &=& 1 + \left( 1 - \gamma_t \right) \cdot \frac{  \mathbf{x}_0^\top \Pid^{-1} \W  \left( \Pf^{-1} + \W^\top \Pid^{-1} \W \right)^{-1} \left( \mathbf{I}_K + \gamma_t \W^\top \Pid^{-1} \W \Pf \right)^{-1} \W^\top \Pid^{-1} \mathbf{x}_0 }{ 
				\mathbf{x}_0^\top \left( \Pid + \W \Pf \W^\top \right)^{-1} \mathbf{x}_0 }.
	\end{eqnarray*}
Restricted to the case when $K=1$,
	\begin{equation*}
		\left( \mathbf{I}_K + \gamma_t \W^\top \Pid^{-1} \W \Pf \right)^{-1} = \left( 1 + \gamma_t \mathbf{w}_1^\top \Pid^{-1} \mathbf{w}_1 \bar{\psi}_{\fund,1} \right)^{-1} = \frac{1}{1+ \gamma_t \cdot \eta_1}.
	\end{equation*}
Consequently,
	\begin{eqnarray*}
		&& \frac{ \mathbf{x}_0^\top \left( \Pid + \gamma_t \W \Pf \W^\top \right)^{-1} \mathbf{x}_0 }{ \mathbf{x}_0^\top \left( \Pid + \W \Pf \W^\top \right)^{-1} \mathbf{x}_0 }
			\\ &=& 1 + \frac{ 1 - \gamma_t }{ 1 + \eta_1 \cdot \gamma_t } \cdot 
				\frac{ \mathbf{x}_0^\top \Pid^{-1} \W  \left( \Pf^{-1} + \W^\top \Pid^{-1} \W \right)^{-1} \W^\top \Pid^{-1} \mathbf{x}_0 }{ 
				\mathbf{x}_0^\top \Pid^{-1} \mathbf{x}_0 - \mathbf{x}_0^\top \Pid^{-1} \W \left( \Pf^{-1} + \W^\top \Pid^{-1} \W \right)^{-1} \W^\top \Pid^{-1} \mathbf{x}_0 }
			\\ &=& 1 + \frac{ 1 - \gamma_t }{ 1 + \eta_1 \cdot \gamma_t } \cdot 
				\left( \frac{ \mathbf{x}_0^\top \Pid^{-1} \mathbf{x}_0 }{
					\mathbf{x}_0^\top \Pid^{-1} \W  \left( \Pf^{-1} + \W^\top \Pid^{-1} \W \right)^{-1} \W^\top \Pid^{-1} \mathbf{x}_0 } -1 \right)^{-1}
			\\ &=& 1 + \frac{ 1 - \gamma_t }{ 1 + \eta_1 \cdot \gamma_t } \cdot 
				\left( \frac{ \mathbf{x}_0^\top \Pid^{-1} \mathbf{x}_0 }{
					\mathbf{x}_0^\top \Pid^{-1} \mathbf{w}_1 \cdot \frac{ \bar{\psi}_{\fund,1} }{ 1 + \bar{\psi}_{\fund,1} \mathbf{w}_1^\top \Pid^{-1} \mathbf{w}_1 } \cdot \mathbf{w}_1^\top \Pid^{-1} \mathbf{x}_0 } -1 \right)^{-1}
			\\ &=& 1 + \frac{ 1 - \gamma_t }{ 1 + \eta_1 \cdot \gamma_t } \cdot 
				\left( \frac{ \mathbf{x}_0^\top \Pid^{-1} \mathbf{x}_0 }{
					\left( \mathbf{w}_1^\top \Pid^{-1} \mathbf{x}_0 \right)^2  } \cdot \frac{ 1 + \eta_1 }{ \bar{\psi}_{\fund,1} } -1 \right)^{-1}.
	\end{eqnarray*}
To simplify notation, define
	\begin{equation*}
		f(x) \triangleq \left( \frac{ \mathbf{x}_0^\top \Pid^{-1} \mathbf{x}_0 }{
					\left( \mathbf{w}_1^\top \Pid^{-1} \mathbf{x}_0 \right)^2  } \cdot \frac{ 1 + \eta_1 }{ \bar{\psi}_{\fund,1} } -1 \right)^{-1}.
	\end{equation*}
Then,
	\begin{eqnarray*}
		\Upsilon(\mathbf{x}_0)
			&=& \sum_{t=1}^T 
				\alpha_t \cdot \left(1 + \theta \cdot (\gamma_t-1) \right)^2 \cdot \left( 1 + \frac{ 1- \gamma_t }{ 1 + \eta_1 \cdot \gamma_t } \cdot f(\mathbf{x}_0) \right).
	\end{eqnarray*}
Note that
	\begin{eqnarray*}
		\sum_{t=1}^T \alpha_t \cdot \left(1 + \theta \cdot (\gamma_t-1) \right)^2
			&=& \sum_{t=1}^T \alpha_t \cdot ( 1 + 2 \theta \cdot (\gamma_t-1) + \theta^2 \cdot (\gamma_t^2-2\gamma_t+1) )
			\\&=& \sum_{t=1}^T \alpha_t + 2 \theta \cdot (\beta_t - \alpha_t) + \theta^2 \cdot \left( \frac{\beta_t^2}{\alpha_t} - 2 \beta_t + \alpha_t \right)
			\\&=& 1 + \theta^2 \cdot \left( \sum_{t=1}^T \frac{\beta_t^2}{\alpha_t} - 1 \right).
	\end{eqnarray*}
As a result,
	\begin{equation*}
		\Upsilon(\mathbf{x}_0)
			= 1 + \theta^2 \cdot \left( \sum_{t=1}^T \frac{\beta_t^2}{\alpha_t} - 1 \right)
				+ \underbrace{ \left( \sum_{t=1}^T \frac{ \alpha_t \cdot \left(1 - \theta \cdot (1-\gamma_t) \right)^2 (1-\gamma_t) }{ 1+ \eta_1 \cdot \gamma_t } \right) }_{\triangleq \Delta } \times f(\mathbf{x}_0).
	\end{equation*}
\qed

\subsection{Proof of Remarks \ref{r-minmax} - \ref{r-individual}}

\noindent
\textbf{Maximum/minimum cost ratio.}
Note that $f(\mathbf{x}_0)$ is a decreasing function of $\frac{ \mathbf{x}_0^\top \Pid^{-1} \mathbf{x}_0 }{ \left( \mathbf{w}_1^\top \Pid^{-1} \mathbf{x}_0 \right)^2 }$, and
	\begin{eqnarray*}
		\min_{\mathbf{x}_0 \in \mathbb{R}^N} \frac{ \mathbf{x}_0^\top \Pid^{-1} \mathbf{x}_0 }{ \left( \mathbf{w}_1^\top \Pid^{-1} \mathbf{x}_0 \right)^2 } 
			&=& \left( \max_{\mathbf{x}_0 \in \mathbb{R}^N}  \frac{ \left( \mathbf{w}_1^\top \Pid^{-1} \mathbf{x}_0 \right)^2 }{ \mathbf{x}_0^\top \Pid^{-1} \mathbf{x}_0 } \right)^{-1}
			= \left( \max_{\mathbf{y} \in \mathbb{R}^N}  \frac{ \left( \mathbf{w}_1^\top \Pid^{-1/2} \mathbf{y} \right)^2 }{ \mathbf{y}^\top \mathbf{y} } \right)^{-1}
			\\&=& \left( \mathbf{w}_1^\top \Pid^{-1} \mathbf{w}_1 \right)^{-1} = \frac{ \bar{\psi}_{\fund,1} }{ \eta_1 }.
	\end{eqnarray*}
The above value is obtained at $\mathbf{x}_0 = \mathbf{w}_1$. Therefore,
	\begin{equation*}
		\max_{\mathbf{x}_0 \in \mathbb{R}^N} f(\mathbf{x}_0) 
			= f(\mathbf{x}_0=\mathbf{w}_1) 
			= \left( \frac{ \bar{\psi}_{\fund,1} }{\eta_1 }\cdot \frac{ 1 + \eta_1 }{ \bar{\psi}_{\fund,1} } - 1 \right)^{-1} 
			= \eta_1.
	\end{equation*}
On the other hand, since $\min_{\mathbf{x}_0 \in \mathbb{R}^N}  \frac{ \left( \mathbf{w}_1^\top \Pid^{-1} \mathbf{x}_0 \right)^2 }{ \mathbf{x}_0^\top \Pid^{-1} \mathbf{x}_0 } = 0$ at $\mathbf{x}_0 = \mathbf{w}_1^\perp$,
	\begin{equation*}
		\min_{\mathbf{x}_0 \in \mathbb{R}^N} f(\mathbf{x}_0) = f(\mathbf{x}_0 = \mathbf{w}_1^\perp) = 0.
	\end{equation*}
Combining these two results, we have that
	\begin{eqnarray*}
		\max_{\mathbf{x}_0 \in \mathbb{R}^N} \Upsilon(\mathbf{x}_0)
			&=& 1 + \theta^2 \cdot \left( \sum_{t=1}^T \frac{\beta_t^2}{\alpha_t} -1 \right)
				+ \max_{\mathbf{x}_0 \in \mathbb{R}^N}\left\{ \Delta \cdot  f(\mathbf{x}_0) \right\}
			= \max\{ \Upsilon_\textup{market}, \Upsilon_\textup{orth} \}
		\\
		\min_{\mathbf{x}_0 \in \mathbb{R}^N} \Upsilon(\mathbf{x}_0)
			&=& 1 + \theta^2 \cdot \left( \sum_{t=1}^T \frac{\beta_t^2}{\alpha_t} -1 \right)
				+ \min_{\mathbf{x}_0 \in \mathbb{R}^N}\left\{ \Delta \cdot  f(\mathbf{x}_0) \right\}
			= \min\{ \Upsilon_\textup{market}, \Upsilon_\textup{orth} \}	,
	\end{eqnarray*}
and trivially, $\Upsilon_\textup{market} \geq \Upsilon_\textup{orth}$ if and only if $\Delta \geq 0$.

\noindent
\textbf{Sign of $\Delta$ with respect to $\theta$.}
Note that $\Delta(\theta)$ is a quadratic function of $\theta$.
It suffices to show that $\Delta(\theta=0) \geq 0$ and $\Delta(\theta=1) \leq 0$.
Note that for an arbitrary function $h(\cdot)$,
	\begin{equation}
	\label{e-inequality}
		\left\{ \begin{array}{ll} 
			\text{if } h(\cdot) \text{ is non-decreasing}, & h(\gamma) \cdot (1-\gamma) \leq h(1) \cdot (1-\gamma), \quad \forall \gamma \\
			\text{if } h(\cdot) \text{ is non-increasing}, & h(\gamma) \cdot (1-\gamma) \geq h(1) \cdot (1-\gamma), \quad \forall \gamma
		\end{array} \right. .
	\end{equation}
In the case of $\theta=0$, by setting $h(\gamma_t) \triangleq \frac{1}{1+\eta_1 \cdot \gamma_t}$ which is a non-increasing function, we can show that
	\begin{equation*}
		\Delta(\theta=0)
		= \sum_{t=1}^T \frac{ \alpha_t \cdot (1-\gamma_t) }{ 1+\eta_1 \cdot \gamma_t }
		\stackrel{ \eqref{e-inequality} }{\geq} \sum_{t=1}^T \frac{ \alpha_t \cdot (1-\gamma_t) }{ 1+\eta_1 }
		= (1+\eta_1)^{-1} \sum_{t=1}^T (\alpha_t - \beta_t) = 0.
	\end{equation*}
In the case of $\theta=1$, by setting $h(\gamma_t) \triangleq \frac{\gamma_t^2}{1+\eta_1 \cdot \gamma_t}$, which is a non-decreasing function,
	\begin{equation*}
		\Delta(\theta=1)
		= \sum_{t=1}^T \frac{ \alpha_t \cdot \gamma_t^2 (1-\gamma_t) }{ 1+\eta_1 \cdot \gamma_t }
		\stackrel{ \eqref{e-inequality} }{\leq} \sum_{t=1}^T \frac{ \alpha_t \cdot (1-\gamma_t) }{ 1 + \eta_1 }
		= 0.
	\end{equation*}
	
\noindent
\textbf{Change of $\Upsilon_\textup{market}$ with respect to $\eta_1$.} Note that
	\begin{eqnarray*}
		\frac{ \partial \Upsilon_\textup{market} }{ \partial \eta_1 }
			&=& \frac{\partial}{ \partial \eta_1} \left( \eta_1\cdot \Delta(\eta_1) \right)
			\\&=& \frac{\partial}{ \partial \eta_1} \left( \sum_{t=1}^T \frac{ \alpha_t \cdot \left(1 - \theta \cdot (1-\gamma_t) \right)^2 (1-\gamma_t) }{ \eta_1^{-1} + \gamma_t } \right)
			\\&=& \sum_{t=1}^T \frac{ \alpha_t \cdot \left(1 - \theta \cdot (1-\gamma_t) \right)^2 (1-\gamma_t) }{ \eta_1^2 \cdot (\eta_1^{-1} + \gamma_t)^2 } 
			\\&=& \frac{ \theta^2 }{ \eta_1^2 } \cdot \sum_{t=1}^T \alpha_t \cdot \left( 1 + \frac{ \theta^{-1} -1 - \eta_1^{-1} }{ \eta_1^{-1} + \gamma_t } \right)^2 (1-\gamma_t).
	\end{eqnarray*}

Set $h(\gamma_t) \triangleq \left( 1 + \frac{ \theta^{-1} -1 - \eta_1^{-1} }{ \eta_1^{-1} + \gamma_t } \right)^2$.
If $\eta_1 \leq \frac{\theta}{1-\theta}$, then $\theta^{-1} - 1 - \eta_1^{-1} \leq 0$ and thus $h(\cdot)$ is non-decreasing. Therefore,
	\begin{eqnarray*}
		\frac{ \partial \Upsilon_\textup{market} }{ \partial \eta_1 }
			&=& \frac{ \theta^2 }{ \eta_1^2 } \cdot \sum_{t=1}^T \alpha_t \cdot \left( 1 + \frac{ \theta^{-1} -1 - \eta_1^{-1} }{ \eta_1^{-1} + \gamma_t } \right)^2 (1-\gamma_t)
			\\ &\stackrel{ \eqref{e-inequality} }{\leq}& \frac{ \theta^2 }{ \eta_1^2 } \cdot \sum_{t=1}^T \alpha_t \cdot \left( 1 + \frac{ \theta^{-1} -1 - \eta_1^{-1} }{ \eta_1^{-1} + 1 } \right)^2 (1-\gamma_t)
			\\ &=& \frac{ \theta^2 }{ \eta_1^2 } \cdot \left( 1 + \frac{ \theta^{-1} -1 - \eta_1^{-1} }{ \eta_1^{-1} + 1 } \right)^2 \cdot \sum_{t=1}^T \alpha_t(1-\gamma_t)
			\\ &=& \frac{ \theta^2 }{ \eta_1^2 } \cdot \left( 1 + \frac{ \theta^{-1} -1 - \eta_1^{-1} }{ \eta_1^{-1} + 1 } \right)^2 \cdot \sum_{t=1}^T (\alpha_t - \beta_t)
			\\&=& 0.
	\end{eqnarray*}
If $\eta_1 \geq \frac{\theta}{1-\theta}$, then $h(\cdot)$ is non-increasing, and thus the sign of the inequality reverses. Therefore,
	\begin{equation*}
		\frac{ \partial \Upsilon_\textup{market} }{ \partial \eta_1 } \leq 0 \quad \text{if } \eta_1 \leq \frac{\theta}{1-\theta}
		, \quad \text{and} \quad
		\frac{ \partial \Upsilon_\textup{market} }{ \partial \eta_1 } \geq 0 \quad \text{if } \eta_1 \geq \frac{\theta}{1-\theta}.
	\end{equation*}
\qed

When $\eta_1 = 0$,
	\begin{equation*}
		\Upsilon_\textup{market}( \eta_1=0)
			= 1 + \theta^2 \cdot \left( \sum_{t=1}^T \frac{\beta_t^2}{\alpha_t} - 1 \right).
	\end{equation*}
Note that $\Upsilon_\textup{market}(\eta_1=0) = \Upsilon_\textup{orth}$.
Since $\Upsilon_\textup{market}(\eta_1)$ is decreasing in $[0, \frac{\theta}{1-\theta}]$, this completes proof of \eqref{e-sign-delta-eta}.
When $\eta_1 = \frac{\theta}{1-\theta}$, since $\frac{ (1-\theta+\theta \cdot \gamma_t)^2 }{ 1 + \eta_1 \cdot \gamma_t } = (1-\theta) \cdot (1-\theta+\theta \cdot \gamma_t)$,
	\begin{eqnarray*}
		\Upsilon_\textup{market}( \eta_1=\frac{\theta}{1-\theta})
			&=& 1 + \theta^2 \cdot \left( \sum_{t=1}^T \frac{\beta_t^2}{\alpha_t} -1 \right) + \frac{\theta}{1-\theta} \cdot (1-\theta) \cdot \sum_{t=1}^T \alpha_t \cdot \left(1 - \theta \cdot (1-\gamma_t) \right) (1-\gamma_t)
			%\\&=& 1 + \theta^2 \cdot \left( \sum_{t=1}^T \frac{\beta_t^2}{\alpha_t} -1 \right) - \theta^2 \sum_{t=1}^T \alpha_t (1-\gamma_t)^2
			\\&=& 1 + \theta^2 \cdot \left( \sum_{t=1}^T \frac{\beta_t^2}{\alpha_t} -1 \right) - \theta^2 \cdot \left( \sum_{t=1}^T \frac{\beta_t^2}{\alpha_t} - 1 \right)
			\\&=& 1.
	\end{eqnarray*}
As $\eta_1 \rightarrow \infty$,
	\begin{eqnarray*}
		\lim_{\eta_1 \rightarrow \infty} \Upsilon_\textup{market}
			&=& 1 + \theta^2 \cdot \left( \sum_{t=1}^T \frac{\beta_t^2}{\alpha_t} -1 \right) + \lim_{\eta_1 \rightarrow \infty} \left( \eta_1 \cdot \Delta(\eta_1) \right)
			\\ &=& 1 + \theta^2 \cdot \left( \sum_{t=1}^T \frac{\beta_t^2}{\alpha_t} -1 \right) + \sum_{t=1}^T \frac{ \alpha_t \cdot \left(1 - \theta \cdot (1-\gamma_t) \right)^2 (1-\gamma_t) }{ \gamma_t }
			\\ &=& 1 + \theta^2 \cdot \left( \sum_{t=1}^T \frac{\beta_t^2}{\alpha_t}-1 \right) + (1-\theta)^2 \cdot \left( \sum_{t=1}^T \frac{\alpha_t^2}{\beta_t} \right) - 1 + 2\theta - \theta^2 \cdot \left( \sum_{t=1}^T \frac{\beta_t^2}{\alpha_t} \right)
			\\ &=& 1 + (1-\theta)^2 \cdot \left( \sum_{t=1}^T \frac{\alpha_t^2}{\beta_t} - 1 \right).
	\end{eqnarray*}
\qed

\noindent
\textbf{Cost ratio of a single stock trading.}
	Note that
	\begin{eqnarray*}
		f(\mathbf{e}_i) 
			&=& \left( \frac{ \mathbf{e}_i^\top \Pid^{-1} \mathbf{e}_i }{ \left( \mathbf{w}_1^\top \Pid^{-1} \mathbf{e}_i \right)^2 } \cdot \frac{ 1 + \eta_1 }{ \bar{\psi}_{\fund,1}} -1 \right)^{-1}
			= \left( \frac{ \bar{\psi}_{\idio,i}^{-1} }{ \left( w_{1i} \cdot \bar{\psi}_{\idio,i}^{-1} \right)^2 } \cdot \frac{1+\eta_1}{\bar{\psi}_{\fund,1}} -1 \right)^{-1}
		\\ &=& \left( \frac{1+\eta_1}{\eta_{1,i}} -1 \right)^{-1}
			= \frac{\eta_{1,i}}{1+\eta_1 - \eta_{1,i}} .
	\end{eqnarray*}
	Also note that
	\begin{equation*}
		\frac{\eta_{1,i}}{1+\eta_1 - \eta_{1,i}} \geq \frac{\eta_{1,j}}{1+\eta_1 - \eta_{1,j}}
		\quad \text{if and only if} \quad
		\frac{w_{1i}^2}{\bar{\psi}_{\idio,i}} \geq \frac{w_{1j}^2}{\bar{\psi}_{\idio,j}} .
	\end{equation*}
	Then, the results immediately follow from \eqref{e-cost-ratio}.
\qed
}
	
\end{document}